\documentclass[journal]{IEEEtran} 
\usepackage{graphicx}
\usepackage{subcaption} 
\graphicspath{ {./Figures/} }
\usepackage{multicol, blindtext} 
\usepackage{array}
\usepackage{amssymb} 
\usepackage{amsmath} 
\usepackage[linesnumbered,ruled,vlined]{algorithm2e}
\usepackage{algpseudocode}
\usepackage{cite}
\usepackage{blindtext}
\usepackage{booktabs} 
\usepackage{float}

\usepackage{tikz,xcolor,hyperref}
\definecolor{lime}{HTML}{A6CE39}
\DeclareRobustCommand{\orcidicon}{%
	\begin{tikzpicture}
	\draw[lime, fill=lime] (0,0) 
	circle [radius=0.16] 
	node[white] {{\fontfamily{qag}\selectfont \tiny ID}};
	\draw[white, fill=white] (-0.0625,0.095) 
	circle [radius=0.007];
	\end{tikzpicture}
	\hspace{-2mm}
}

\foreach \x in {A, ..., Z}{%
	\expandafter\xdef\csname orcid\x\endcsname{\noexpand\href{https://orcid.org/\csname orcidauthor\x\endcsname}{\noexpand\orcidicon}}
}

\begin{document}
	
    \title{Design and Development of a Scalable and Energy-Efficient Localization Framework Leveraging LoRa Ranging-Capable Transceivers}
	
	\author{
		Hasan~Albinsaid\orcidA{},~\IEEEmembership{Student Member,~IEEE,}
		Bodhibrata~Mukhopadhyay\orcidB{},~\IEEEmembership{Member,~IEEE,}
		and~Mohamed-Slim~Alouini\orcidC{},~\IEEEmembership{Fellow,~IEEE}
		\thanks{Hasan Albinsaid and Mohamed-Slim Alouini are with the Division of Computer, Electrical and Mathematical Science and Engineering, King Abdullah University of Science and Technology, Thuwal 23955-6900, Saudi Arabia. Bodhibrata Mukhopadhyay was with CEMSE, KAUST, Thuwal, 23955-6900, Kingdom of Saudi Arabia. He is now with the Department of Electronics and Communication, Indian Institute of Technology Roorkee, Uttarakhand, 247667, India  (email: hasan.albinsaid@kaust.edu.sa; slim.alouini@kaust.edu.sa, bodhibrata@ece.iitr.ac.in)}
	}
	
	\maketitle
	
	\begin{abstract}
		Precise and energy-efficient localization is a critical requirement in many Internet of Things (IoT) applications, particularly in large-scale deployments such as asset tagging, agriculture, and smart cities, where long battery life and cost-effectiveness are crucial. The Semtech SX1280 LoRa transceiver presents a promising solution for IoT localization. It combines low cost, low power, and precise ranging capability over distances of up to 1 km. However, the ranging process requires two devices to be simultaneously active, one initiating the ranging request and the other responding to it, which can lead to significant energy expenditure if not properly managed. Despite the transceiver’s excellent performance, no existing system-level framework effectively manages sleep-wake coordination and role assignment needed for energy-efficient operation.  This paper presents a coordination framework that significantly reduces power consumption while maintaining the inherent precise ranging capability of the chip. The framework schedules short, synchronized wake-up windows between the \textit{initiator} and the \textit{responder}, allowing devices to remain in deep sleep for most of their duty cycle. This scheduling strategy minimizes reliance on precise continuous timing and mitigates drift in low-cost oscillators.  To validate the framework, we designed and developed custom nodes that are compliant with the framework’s protocol. Experimental results show that the proposed approach allows a node to stay in ultra-low power mode and wake periodically to check for instructions. The node can remain in standby mode for up to nine months on a single coin cell battery and can perform ranging operations on demand in near real-time, all while maintaining a localization accuracy within five meters.
	\end{abstract}
	
	
	\section{Introduction}
	\IEEEPARstart{L}{ocalization} plays a crucial role in the context of Internet of Things (IoT) devices. These devices are often spread across a specific area to monitor and measure various parameters of interest. The data these devices collect is meaningful only when linked with their accurate locations. Therefore, estimating the locations of these devices is a key requirement for some practical applications such as logistics, agriculture, and smart cities\cite{Panwar2022_localiz, Wang2025LocalizationFGCS, Katakwar2025IndoorTracking}. These IoT devices, which are small, low-cost, and low-power nodes, are commonly deployed in large numbers over a region of interest with limited control over their location in space \cite{Salari2018Distributed, Zhao2023Underwater}. For example, they might be dispersed in an environment for monitoring, sensing, or tagging purposes. Equipping each device with a Global Positioning System (GPS) receiver would significantly increase the network costs, power consumption, and limit applicability in some cases, such as in mining activities \cite{Habibzadeh2017SmartCity, Ahmed2022LoRa, zare2021applications}. To keep implementation costs low, some works suggest only a small fraction of devices are equipped with GPS receivers (known as anchors), while the remaining devices (known as targets) determine their locations using a localization scheme that leverages the known anchor locations \cite{Nguyen2019LEMON}. 
	
	While localization schemes can utilize various types of measurements such as range, angle, or signal strength, range-based methods remain the most widely adopted due to their higher accuracy and broader applicability across diverse environments~\cite{slavisa2018}. There are several ways of obtaining range measurements based on radio signals, such as time of arrival (ToA), Time Difference of Arrival (TDoA), round-trip time~\cite{Hadir2021Accurate, Han2016Survey, Sesyuk2022A}. 
	Time-of-Arrival (ToA) has emerged as a popular method for radio signal-based localization, driven by its ability to deliver high accuracy in distance estimation and positioning~\cite{ravindra2014time,  mekonnen2014robust}. However, it also introduces several practical challenges; it requires precise clock synchronization between devices, which can be difficult to achieve in practice, and the method is sensitive to signal propagation delays, which can be affected by various environmental factors and lead to inaccuracies in location estimation~\cite{Wu2022AConvex, vonTschirschnitz2019}.
	
	The time synchronization challenge in ToA-based systems can be addressed using the Two-Way ToA (TW-ToA) method, which eliminates the need for precise clock synchronization. This approach has been implemented in Semtech’s SX1280 \cite{semtechsx1280}, a LoRa transceiver that supports both standard Chirp Spread Spectrum (CSS) communication and TW-ToA ranging, making it well-suited for both communication and ranging applications. In the TW-ToA technique, ranging between two nodes involves a bidirectional exchange of packets, which typically completes in just a few milliseconds. Similar to standard LoRa communication, it also requires the transceivers to be configured on the same operational parameters: center frequency, Spreading Factor (SF), code rate, and bandwidth~\cite{semtechsx1280, Sezana2024F}. In TW-ToA, one node is designated as the initiator and begins the process by transmitting a ranging request packet. Upon receiving this packet, the responder node replies by sending a response packet back to the initiator. This interaction enables the initiator to estimate the round-trip time and, consequently, compute the accurate distance between the two nodes. Given its ability, LoRa ranging-capable transceivers are an appealing choice for wireless network-based localization due to their low power consumption, high accuracy, wide coverage, and relatively low cost \cite{muller2021outdoor, ghodhbane2024pressure}. Despite its power consumption being lower than that of many alternative technologies, it still poses a challenge for energy-constrained IoT devices. Therefore, further optimization is essential to improve overall energy efficiency.
	
	To enhance energy efficiency in TW-ToA-based localization, an effective approach is to keep nodes in sleep mode and wake them only when a ranging operation is needed. Given that the duration of the ranging exchange is on the order of milliseconds, implementing such a strategy poses a significant challenge: how to coordinate ranging among nodes that are asleep most of the time. A widely adopted solution involves using Real-Time Clocks (RTCs) to schedule periodic wake-up windows \cite{callebaut2021art, mani2018architecture, ahmed2025investigating, kozlowski2019energy, kazdaridis2020nano}. However, this method is constrained by the inherent limitations of low-cost quartz crystal oscillators commonly used in such devices, which are prone to substantial drift, often exceeding seconds within just a few days \cite{ClockSources2018, capriglione2016analysis}. Consequently, maintaining precise coordination across distributed nodes becomes impractical, thereby complicating the orchestration of reliable and energy-efficient ranging schedules.

	While LoRa ranging-capable transceivers technology offers significant benefits for both ranging and communication, the critical challenge of coordinating these interactions remains largely unaddressed. Despite these capabilities, no framework has yet been developed that efficiently orchestrates ranging operations among LoRa ranging-capable devices in an energy-efficient manner. In this paper, we address this gap:
	
	\begin{enumerate}  
		\item We propose a novel localization framework that coordinates nodes to orchestrate sleep, wake-up, transmission, and reception operations associated with the ranging operation. This coordinated approach significantly reduces overall nodes' energy consumption while preserving the inherent ranging accuracy and communication capabilities of LoRa ranging-capable transceivers. The framework facilitates precise wake-up of specific node pairs through a countdown-based scheduling strategy, making nodes resilient to clock drift in low-cost IoT devices.

        \item 
        We implement the proposed framework on a microcontroller + SX1280 testbed and validate it through real-world experiments, demonstrating feasibility on practical low-power hardware.
		
		\item We make the complete implementation of the proposed framework publicly available: the network server, hardware design, and firmware.\footnote{The repository hosted at: \url{https://github.com/hasanabs/rangingstack}}
		
	\end{enumerate}
	
	The rest of this paper is organized as follows: Section II discusses related work in improving the energy efficiency in localization, Section III provides an overview of the characteristics of the LoRa ranging-capable transceivers chip for ranging, serving as a technical background. Sections IV and V present the proposed framework design, detailing how orchestrated interaction minimizes power consumption while maintaining localization precision. Section VI describes the implementation, experimental setup, and evaluation of the framework’s performance through real-world tests. Section VII discusses potential extensions of the current solution. Finally, Section VIII concludes the paper with a summary of findings and potential applications.

    \section{Related Work}

	Energy-efficient localization has long been a focus in wireless IoT systems, with various technologies offering different trade-offs in accuracy, power consumption, and deployment complexity \cite{tarrio2013energy, cox2020high}. Prior work outside the LoRa ecosystem has explored hybrid methods combining low-power radios with high-precision ranging technologies. A prominent example is Apple AirTag, which uses Bluetooth Low Energy (BLE) for low-power coarse location prediction and Ultra Wideband (UWB) for precise location prediction. In its default mode, the tag broadcasts BLE beacons to conserve energy. When an Apple device (iPhone 11 or newer) detects the beacon, it forwards the information to Apple’s server so that the owner can access the approximate location of the tag, which is inferred from the Received Signal Strength Indicator (RSSI). When the owner requests the precise location, the Apple device relays this request to the tag, which then activates UWB to achieve sub-meter accuracy \cite{sung2023uwb}. This design, however, has notable limitations: BLE is restricted to less than 50 meters in open areas, UWB consumes considerable power and is limited to less than 30 meters in open areas, and both require proximity to Apple devices to serve as intermediaries. Consequently, AirTag is unsuitable for wide-area or remote IoT deployments~\cite{AppleAirTag2021}.
    
	Other studies have explored energy savings through Wake-up Receivers (WuR)~\cite{Polonelli2021WUR, cortesi2025wakeloc}, which allow nodes to remain in deep sleep until activated by a wake-up signal. This greatly reduces idle listening energy but poses a challenge for localization. Specifically, the wake-up trigger must originate near the node, implying prior knowledge of its location. Furthermore, WuR-based designs require additional hardware, increasing system complexity and cost. Similarly, GPS duty cycling combined with short-range radio ranging has been used to reduce energy consumption~\cite{ACM2013_GPS_DutyCycling}. In this method, GPS modules are activated intermittently, while low-power radios provide interim positioning. Although energy use is reduced by approximately one-third, GPS remains a major energy bottleneck due to its high acquisition cost and startup delays after sleeping.
    
	Another relevant direction involves ultra-low-power localization is ``radio landmarks'' which is integrated into infrastructure devices like smoke detectors~\cite{Simon2015RadioLandmarks}. These systems use a mix of RSSI and Inertial Measuring Unit (IMU) data to estimate indoor position while consuming only microwatts in standby. However, they depend on dense deployment of anchors and are more suitable for controlled indoor environments rather than large-scale outdoor IoT networks. Moreover, there is no conclusive evidence has been provided to demonstrate their energy efficiency in real deployments.
    
	Within the LoRa domain, a substantial body of research has focused on RSSI-based localization using sub-GHz LoRa transmissions~\cite{Lam2021RSSIbased, vazquez2020experimental, Han2016Survey}. These methods estimate distance from the received signal strength of standard communication packets, often using empirical models or fingerprinting. While inherently low-power, RSSI-based approaches suffer from poor ranging accuracy, typically between 10 to 100 meters, and are highly sensitive to environmental noise, interference, and multipath propagation.
    
	Although LoRa was originally developed solely for low-power and long-range communication, its CSS modulation shares similarities with radar signals, making it possible to extract distance information from propagation delays. Building on this insight, the Semtech SX1280 chip was introduced, operating at 2.4 GHz and uniquely supporting both communication and TW-ToA ranging\cite{Rander2020Ranging, muller2021outdoor, Janssen2020LoRa, Gwendoline2023Oppor}. Existing studies primarily evaluate the chip’s ranging performance under different conditions. For example, author \cite{Janssen2020LoRa} analyzes accuracy across antenna orientations and spreading factors, while \cite{muller2021outdoor} demonstrates sub-5 meter accuracy using TW-ToA-based multilateration in open outdoor settings. However, these efforts focus exclusively on the ranging capability or hardware performance, without addressing energy-aware scheduling, coordination, or protocol-level improvements.
    
	To the best of our knowledge, no existing research proposes a framework-level protocol for LoRa ranging-capable transceivers that enables accurate TW-ToA ranging while keeping nodes in deep sleep for most of their duty cycle. Our work addresses this gap by introducing a countdown-based coordination protocol that facilitates short synchronized wake-ups, tolerates clock drift in low-cost devices, and eliminates unnecessary energy consumption. Unlike prior efforts that remain limited to simulations or hardware benchmarking, we implement and validate our framework through real-world deployment. The system achieves months-long battery life on a CR2032 coin-cell battery, without sacrificing the inherent high ranging accuracy of the TW-ToA method. This demonstrates that our framework, when implemented on low-power LoRa ranging-capable hardware, uniquely achieves both the inherent high ranging precision of TW-ToA and months-long battery life through energy-aware scheduling.

    \section{Characteristics of LoRa Ranging-Capable Transceivers}

    LoRa technology across all frequency bands (e.g., 433~MHz, 865~MHz, 868~MHz, 915~MHz, and 2.4~GHz) relies on CSS modulation, where each symbol represents $SF$ bits by using one of $2^{SF}$ possible cyclic time shifts of the chirp. This property not only supports robust long-range communication but also provides the foundation for Time-of-Flight (ToF) based ranging. 

    Among currently available transceivers, the Semtech SX1280 is the first commercial device to expose ranging functionality in addition to communication. Operating in the 2.4~GHz ISM band, it offers wider bandwidth options than sub-GHz LoRa transceivers, with configurable bandwidths of 203~kHz, 406~kHz, 812~kHz, and 1625~kHz, and spreading factors ranging from 5 to 12~\cite{Gwendoline2023Oppor, polak2020performance, Rander2020Ranging, Janssen2020LoRa}. For example, at an SF of 10 with a 1625~kHz bandwidth, the SX1280 achieves a receiver sensitivity of $-120$~dBm, corresponding to a maximum coupling loss of 132.5~dB when transmitting at 12.5~dBm \cite{semtechsx1280}. This supports communication and ranging over several kilometers in rural or line-of-sight scenarios. Although the SX1280 currently dominates the literature on LoRa ranging, the same principles can be extended to other bands should future LoRa transceivers expose similar ranging features.
	
	\begin{figure}[!t]
		\centering
		\includegraphics[width=1 \columnwidth, trim={.72cm 0 0.22cm 0}, clip]{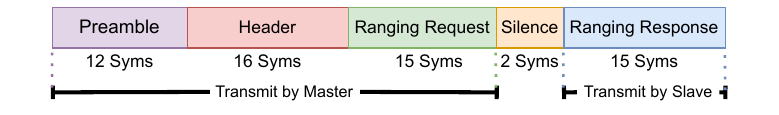}
		\caption{The ranging frame of SX1280 ranging engine}
		\label{fig:protocolSX}
	\end{figure}

    In the physical frame of SX1280, the ranging engine employs a two-way ranging procedure with half-duplex communication systems. This protocol includes a preamble, header, ranging request, and ranging response, as illustrated in Fig.~\ref{fig:protocolSX}. Compared to one-way ranging methods, this two-way approach eliminates the need for clock synchronization between the nodes. However, to maintain ranging accuracy, the frequency offset between the oscillators of the two nodes must remain minimal. In practice, the SX1280 mitigates this by monitoring the duration of the ranging response. Any deviation from the expected ranging response reveals instantaneous oscillator error, which can be compensated in the ToF calculation. This improves the per-exchange accuracy, but does not address long-term RTC drift across multiple ranging sessions.

    \begin{figure}[!b]
        \centering
        \includegraphics[width=0.56 \columnwidth, trim={.72cm 1cm 0.22cm 1cm}, clip]{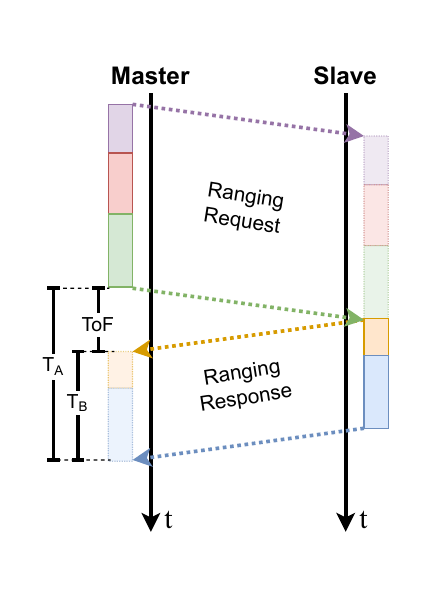}
        \caption{Timing diagram of the SX1280 ToF ranging frame}
        \label{fig:protocolSX_time}
    \end{figure}
    
    Fig.~\ref{fig:protocolSX_time} illustrates the operational mechanism of ToF measurement in the SX1280 transceiver.
    Consider two nodes, referred to as \textit{master} and \textit{slave}, both preconfigured with identical configuration parameters (e.g., SF, bandwidth, center frequency). The master initiates the process by sending a ranging request to the slave’s address, ensuring that the correct node receives the request among many possible active nodes in the network. Once the entire request packet is sent, the master starts its timer. Once the slave fully receives the request packet, it remains silent for a predefined switching delay $T_D$ before responding. This delay accounts for internal processing time and is defined as $T_D = 2 \cdot T_s$, where $T_s = 2^{\text{SF}} / \text{BW}$ is the duration of one chirp symbol in the CSS modulation used by LoRa. After this delay, the slave sends a response packet back to the master, completing the ranging exchange. When the master receives the entire ranging response packet, it stops its timer. The measured duration, denoted as $T_A$, is then used to calculate the ToF $\left(ToF=T_A-T_B\right)$. Since both the master and the slave share the same configuration, the value of $T_B$ can be determined on the master’s side. Given that the value of $T_B$ is the sum of the silence period and the ranging response duration, which consists of 17 symbols $\left(T_B=17\cdot T_s\right)$. Here it is useful to explicitly remind that this is equivalent to a radar-style time-of-flight, where the signal delay is proportional to distance ($d = c \cdot \Delta t$ with $c$ being the speed of light). This connection makes clear why sub-microsecond timing errors translate directly into meter-scale ranging errors. Finally, the one-way distance between the master and slave is calculated using:
    
    \begin{figure}[!b]
        \centering
        \includegraphics[width=1.0 \columnwidth]{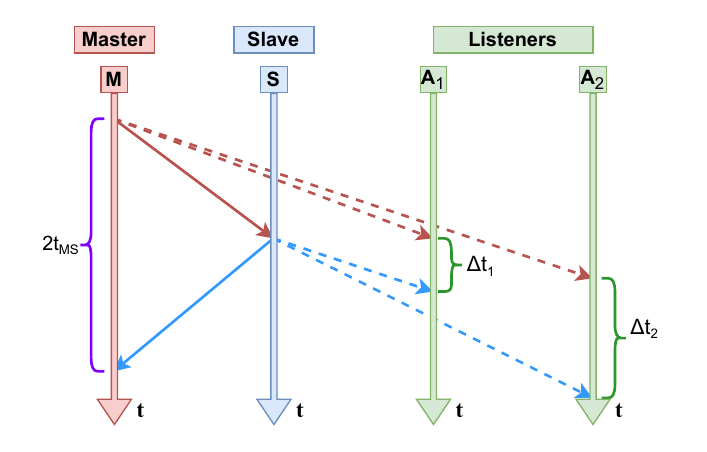}
        \caption{Timing diagram of the point-to-point ranging between a master and a slave and their passive listeners.}
        \label{fig:workingofAr}
    \end{figure}
    
    \begin{IEEEeqnarray}{c}
        d = c \cdot \frac{ToF}{2} = c \cdot \frac{(T_A-T_B)}{2}.
        \label{eq:tof}
    \end{IEEEeqnarray}
    
    However, since the master calculates the value of \(T_B\) based on its own CSS symbol duration \(T_s\), any oscillator frequency offset can lead to discrepancies in $T_s$, thereby introducing errors in the distance measurement computed using~\eqref{eq:tof}. Let $f_M$ and $f_S$ denote the oscillator frequencies of the master and slave, respectively, and the relative oscillator offset is defined as \(\delta=(f_M-f_S)/f_0\), with $f_0$ being the nominal frequency. We can approximate the measurement drift in ranging by:
    
    \begin{IEEEeqnarray}{c}
        \Delta d = c\cdot\frac{\delta T_B}{2}.
        \label{eq:tof_error}
    \end{IEEEeqnarray}
    
    Based on that, we can obtain the corrected distance using
    \begin{IEEEeqnarray}{c}
        d_{\text{corr}} = c\cdot\frac{(T_A-(1+\delta)\cdot T_B)}{2}.
        \label{eq:d_corrected}
    \end{IEEEeqnarray}
    
    Accordingly, compensating for oscillator drift can mitigate ranging errors. This is exactly where the SX1280’s built-in correction comes in: by comparing the measured vs. expected $T_B$, the chip estimates $\delta$ and applies compensation as in Eq.~\eqref{eq:d_corrected}. This compensates instantaneous oscillator drift during a single ranging exchange. However, note that $\delta$ may vary dynamically (e.g., due to temperature), and cannot serve as a stable long-term clock reference for scheduling across days.

	Besides the direct ranging method discussed earlier, the SX1280 chip also supports passive ranging (also known as Advanced Ranging (AR)), where multiple nodes can simultaneously receive ranging packets \cite{Semtech2024Ranging}. This approach extends the traditional point-to-point ranging concept, as illustrated in Fig.~\ref{fig:workingofAr}. In this illustration, passive listeners measure the time difference between the master's ranging request and the slave's response. This time difference, denoted as $\Delta t_j$, for the $j^{\text{th}}$ listener $(j\in\{1,2,\dots, J\})$, is given by:
	\begin{equation}
		\begin{aligned}
			\Delta t_{j} &= (t_{MS}+t_{SA_j})\;-\;t_{MA_j} 
		\end{aligned}
		\label{eq:delta_t_j}
	\end{equation}
	where \(t_{MS}\) is the ToF between the master and the slave, \(t_{SA_j}\) is the ToF between the slave and the $j^{\text{th}}$ listener, and \(t_{MA_j}\) is the ToF between the master and the $j^{\text{th}}$ listener. 
	
	A passive listener node can measure the relative time difference between the ranging request and response packets, which is directly proportional to the difference in its distances to the master and the slave. Geometrically, this implies that the listener lies on a hyperboloid surface with the master and slave nodes located as its foci. Such passive ranging is particularly valuable in localization and tracking scenarios, where multiple observer nodes estimate their spatial relationships based solely on overheard ranging messages.
	
	\section{Proposed Framework Design}
	\begin{figure*}[!t]
		\centering
		\includegraphics[width=1.72 \columnwidth, trim={0.72cm 0 0.8cm 0}, clip]{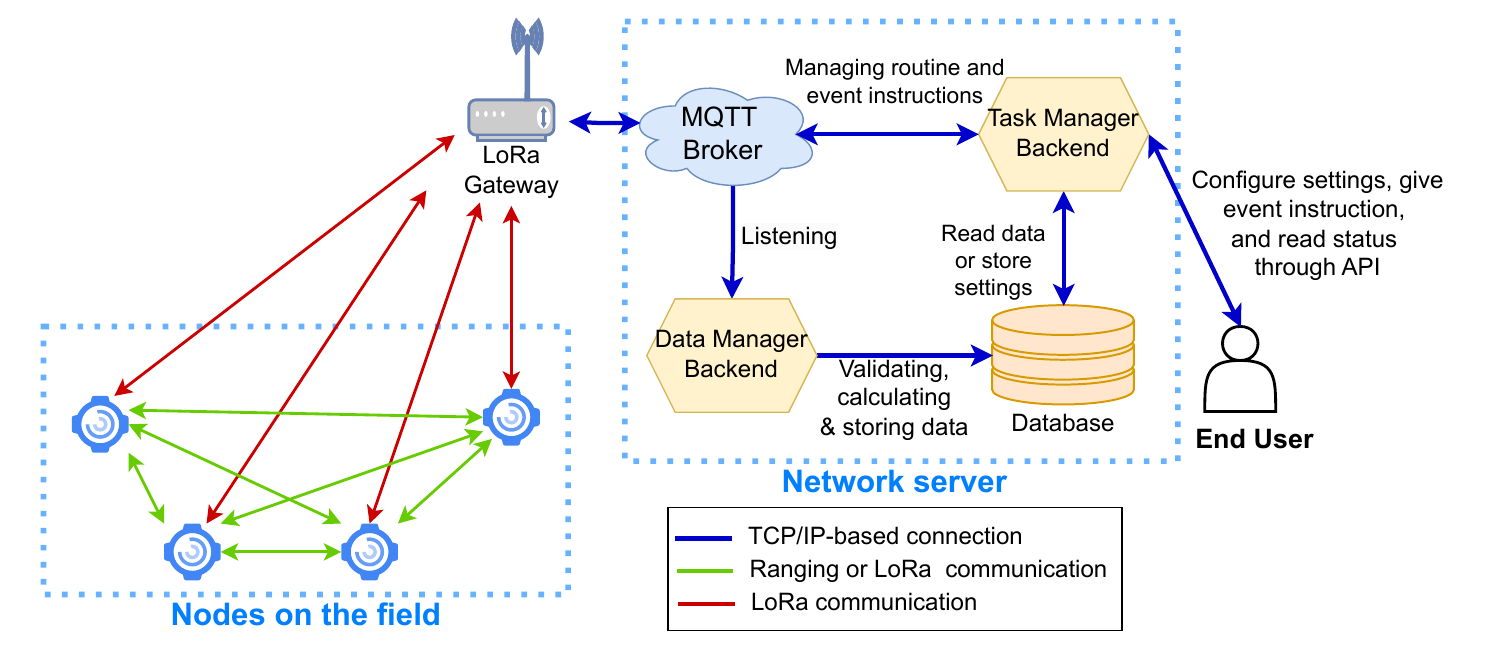}
		\caption{Architecture of the Proposed Localization Framework Using LoRa Ranging-capable Transceiver}
		\label{fig:framework}
	\end{figure*}
	
	We developed a novel localization framework that uses LoRa ranging-capable transceivers for both communication and ranging. The proposed framework incorporates an energy-aware scheduling mechanism that precisely instructs nodes to wake up only for designated tasks, allowing them to remain in deep sleep for the majority of their duty cycle. The framework comprises three main components: wireless nodes, a Gateway (GW), and a Network Server (NS), as illustrated in Fig.~\ref{fig:framework}.
	
	The first component is the wireless node, which is equipped with a LoRa ranging-capable transceiver. Nodes can be configured either as anchors, which are reference nodes with known locations, or as targets, whose location is to be estimated through the localization process. To conserve power, nodes are designed to remain in sleep mode most of the time, waking only for a few milliseconds at configurable intervals to perform an instruction check. During this process, a node briefly powers up to contact the gateway and inquire whether any instructions are queued for it, such as ranging tasks or configuration updates. The second component is GW, which integrates a LoRa ranging-capable transceiver with TCP/IP-based connectivity. The gateway communicates with the NS using the MQTT protocol to coordinate tasks and relay data. Finally, the NS acts as an orchestrator to schedule interactions between nodes and processes the ranging data to compute the targets' locations.
	
	Scheduling is a key challenge for low-power, low-cost nodes~\cite{Mahmoud2019, ji2022intelligent}. This framework addresses these challenges through orchestrated interactions. Since low-cost oscillators are prone to drift, maintaining absolute time is not feasible~\cite{Mahmoud2018FADS}. Instead, nodes coordinate their activities based on instructions received from the GW, which include a countdown timer specifying when each task should be executed.  The GW acts as a relay, forwarding this information to the nodes from the NS. Here, the countdown is stored in the NS as a variable corresponding to its instruction, which decreases over time.
	\begin{figure}[!b]
		\centering
		\begin{subfigure}{0.30\textwidth}
			\centering
			\includegraphics[width=\linewidth]{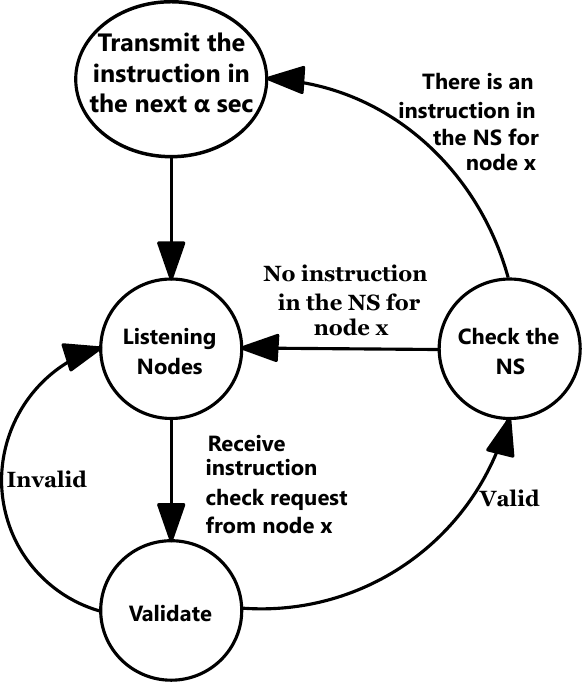}
			\caption{}
			\label{fig:sub1GW}
		\end{subfigure}
		\hfill
		\begin{subfigure}{0.29\textwidth}
			\centering
			\includegraphics[width=\linewidth]{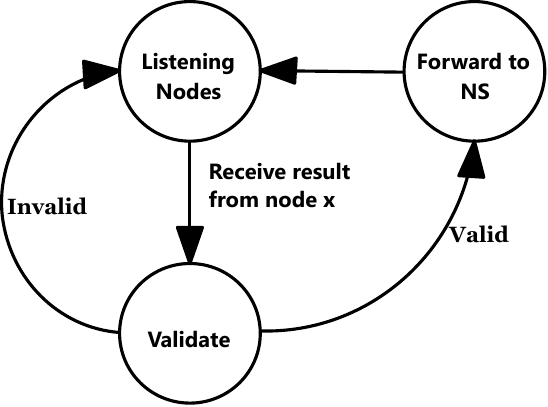}
			\caption{}
			\label{fig:sub2GW}
		\end{subfigure}
		\caption{State diagram of a GW where (a) is an instruction check handler, and (b) is a result handler.}
		\label{fig:StateDiagramGW}
	\end{figure}
	
	Fig.~\ref{fig:StateDiagramGW} illustrates the state diagram of the GW. The GW operates through two concurrent routines: (1) the instruction check handler and (2) the result handler. Both routines run concurrently without blocking each other. 
	In the instruction check handler, shown in Fig.~\ref{fig:StateDiagramGW}(a), a GW receives an instruction check request from a node, and it first validates the data before forwarding it to the NS. If the NS says no instruction is available for that node, the GW does not send a response. Otherwise, the GW queues the instruction and transmits it to the node after a predetermined delay (e.g., $\alpha$ seconds). This delay allows the node to enter sleep mode while waiting, ensuring that even if network latency introduces a delay of a few hundred milliseconds, it remains within an acceptable range. Next is the result handler, which is shown in Fig.~\ref{fig:StateDiagramGW}(b). For example, when a node transmits ranging results, the GW validates the data and forwards them to the NS for further processing. This bridging approach ensures efficient power management while maintaining reliable communication between nodes and the NS.
	
	\begin{figure}[ht!]
		\centering
		\begin{subfigure}{0.47\textwidth}
			\centering
			\includegraphics[width=\linewidth]{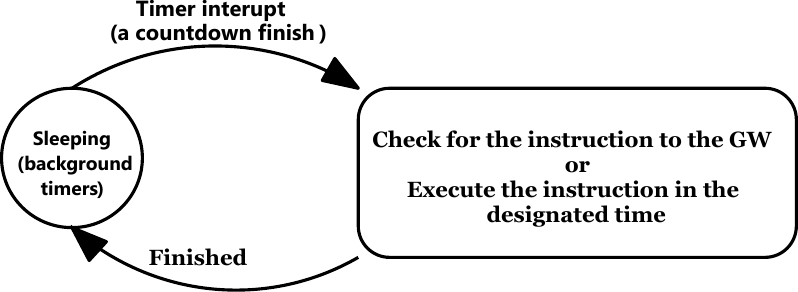}
			\caption{}
			\label{fig:sub1Node}
		\end{subfigure}
		\hfill
		\begin{subfigure}{0.5\textwidth}
			\centering
			\includegraphics[width=\linewidth]{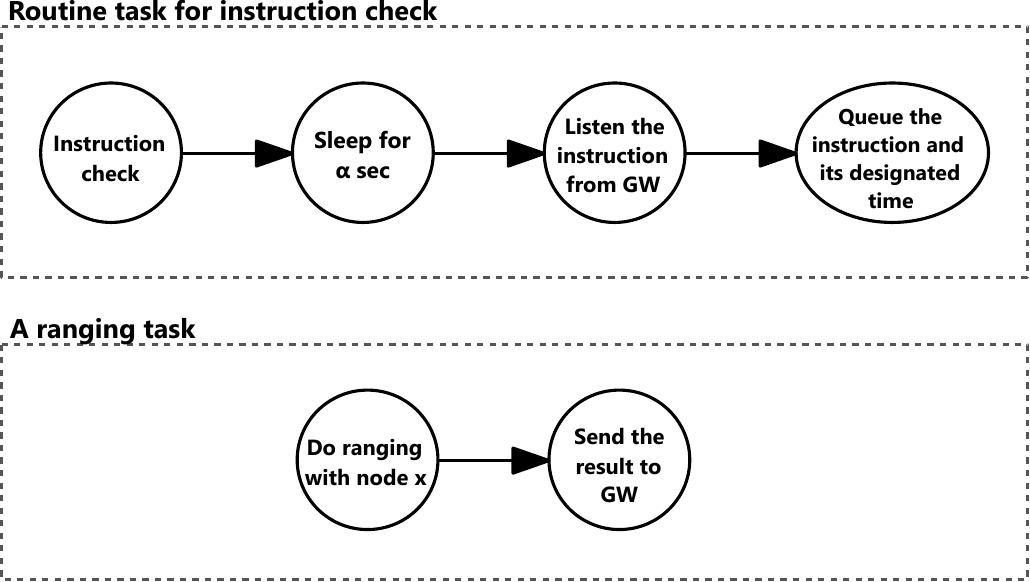}
			\caption{}
			\label{fig:sub2Node}
		\end{subfigure}
		\caption{State diagram of a node: (a) general operational flow, and (b) examples of routine and non-routine tasks.}
		\label{fig:StateDiagramNode}
	\end{figure}
	
	Next, we discuss the state transition diagram of a node as illustrated in Fig.~\ref{fig:StateDiagramNode}. The node runs a Real-Time Operating System (RTOS), which enables multitasking by allowing multiple tasks to be scheduled and managed efficiently by the operating system’s kernel. As shown in Fig.~\ref{fig:StateDiagramNode}(a), the node primarily remains in sleep mode and wakes only in response to scheduled events. These events include periodic instruction checks or task execution, which is already being scheduled. Both operations are triggered by a background timer interrupt, which operates at a low frequency to minimize power consumption. Upon task completion, the node returns to sleep mode, ensuring minimal energy usage. Fig.~\ref {fig:StateDiagramNode}(b) presents two example tasks that the node can execute. The first is a routine instruction-check task, which occurs at configurable intervals. During this task, the node wakes, listens for instructions from the GW, and queues any received commands with their designated execution times in a countdown form. The second is a non-routine task, a ranging task, which is initiated on demand to measure the distance between two nodes. In this task, the node performs a ranging operation with a specified peer node and transmits the result to the GW.
	
\begin{figure}[!b]
		\centering
		\includegraphics[width=1 \columnwidth]{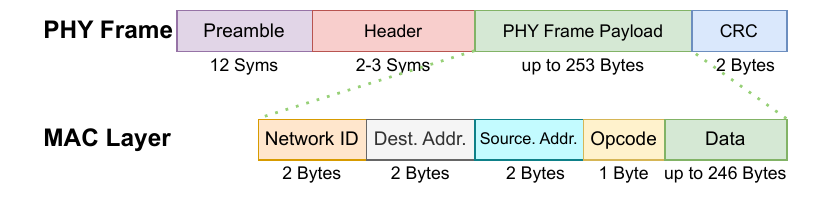}
		\caption{Communication frame structure used in the proposed framework}
		\label{fig:COmmunFrame}
	\end{figure}

	Following the explanation of the framework’s state transition diagrams, the next important aspect is the communication mechanism between the nodes and the gateway. In this framework, we designed a dedicated communication protocol as depicted in Fig.~\ref{fig:COmmunFrame}.
	At the Physical (PHY) frame, each transmitted packet is composed of four main parts:
	\begin{itemize}
		\item Preamble (12 symbols): Used for synchronization between the transmitter and receiver.
		\item Header (2-3 Symbols): Encodes essential transmission parameters, including the size of the PHY payload, the code rate selected for Forward Error Correction (FEC), and a CRC flag indicating whether the header includes CRC protection.
		\item PHY Frame Payload (dynamic, up to 253 bytes): Carries the actual data content, encapsulating the higher-layer protocol (MAC layer).
		\item CRC (2 bytes): Provides a checksum for detecting errors in the transmitted packet.
	\end{itemize}
	Encapsulated within the PHY frame payload is the Medium Access Control (MAC) layer protocol structure. The MAC frame is organized as follows:
	\begin{itemize}
		\item Network ID (2 bytes): Identifies the network to which the packet belongs, enabling network segregation in a case where a cluster network is needed.
		\item Destination Address (2 bytes): Specifies the target node or gateway for the packet.
		\item Source Address (2 bytes): Identifies the originating node.
		\item Opcode (1 byte): Defines the type of operation or message, such as instruction check, ranging task assignment, status update, or configuration update command.
		\item Data (dynamic, up to 246 bytes): Contains the operation-specific payload, with its structure and interpretation strictly defined by the associated opcode. Although the Data field is flexible in size to support varying application needs, it must conform to the format specified by the opcode to ensure protocol consistency.
	\end{itemize}
	With this standardized packet structure, each node can correctly interpret and process messages from other nodes and the gateway, enabling seamless and consistent communication across the network.
	
	\begin{figure}[!t]
		\centering
		\includegraphics[width=0.85 \columnwidth]{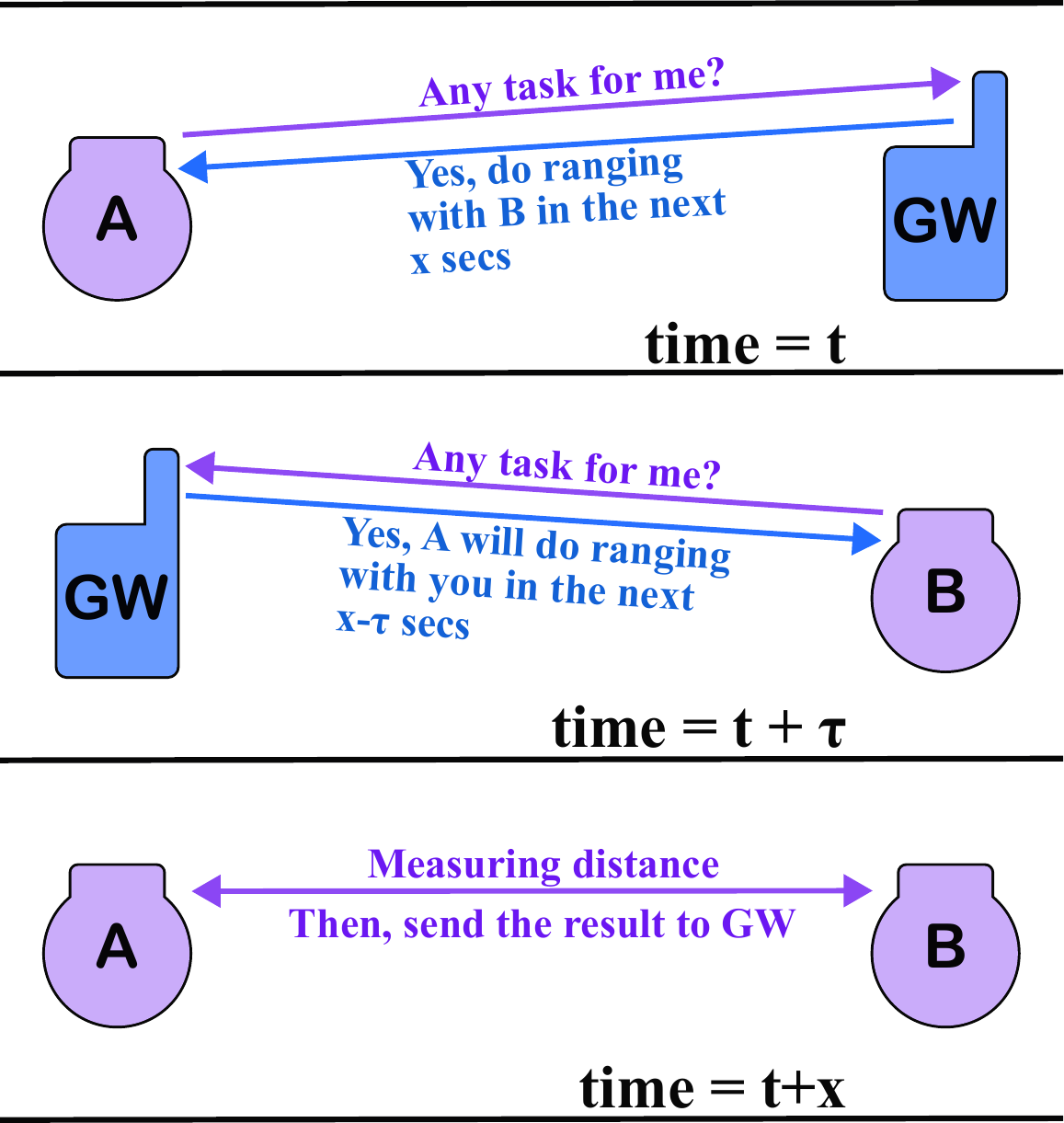}
		\caption{Orchestration example for doing point-to-point (PtP) ranging between node A and node B}
		\label{fig:frameworkIlustration}
	\end{figure}
	
	To help us figure out how the orchestration process works, Fig.~\ref{fig:frameworkIlustration} shows the illustration of the ranging exchange between Node A and Node B. Assume that a ranging task is queued in the NS for these two nodes and the task’s countdown is set to be greater than the maximum of the remaining time for instruction check of both Node A and Node B. When Node A wakes up, it queries the GW for any assigned tasks. The GW retrieves the relevant task from the NS and instructs Node A to initiate ranging with Node B in the next $x$ seconds. Similarly, when Node B wakes up (at time $t+\tau$), it receives a task to participate in the ranging with A in the next $x-\tau$ seconds. At the scheduled time, Nodes A and B wake up and perform ranging, and Node A sends the results to the GW, which forwards the data to the NS for further computation.

    Signal collisions are a known challenge in LoRa-based communication. In our framework, to address this, instructions are not queued solely as static payloads, as is typical in LoRaWAN, but rather as static instructions paired with a dynamic countdown variable. This design ensures that even if a collision occurs during an instruction check and the node must retry the request at a later time (randomly delayed within a bounded window to mitigate further collisions),  the correct \emph{remaining} countdown value is still received. As a result, the timing of time-sensitive tasks, such as ranging, remains consistent and accurate despite the potential for retransmissions.

    \begin{figure}[!t]
        \centering
        \begin{subfigure}[b]{1.0\columnwidth}
            \centering
            \includegraphics[width=\linewidth, trim={2.1em 1em 2em 1em}, clip]{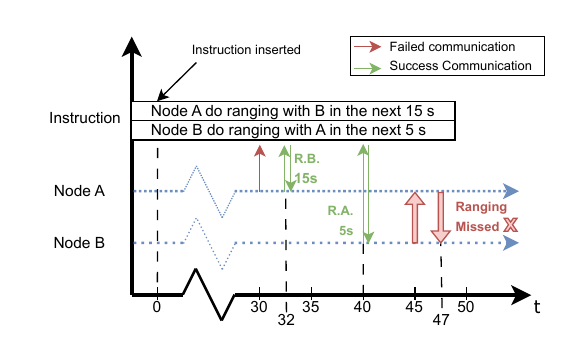}
            \caption{}
        \end{subfigure}
        \hfill
        \begin{subfigure}[b]{1.0\columnwidth}
            \centering
            \includegraphics[width=\linewidth, trim={2.1em 1em 2em 1em}, clip]{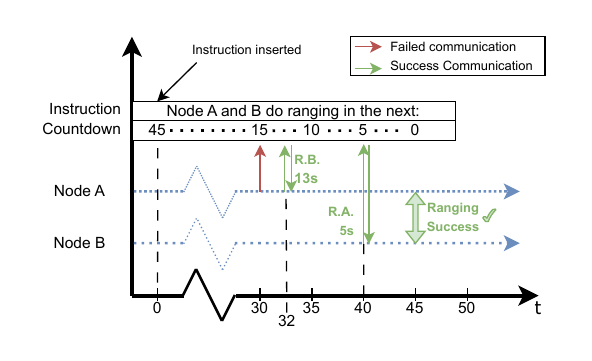}
            \caption{}
        \end{subfigure}
        
        \caption{Timing diagram illustration of using  (a) LoRaWAN approach, and (b) Proposed framework approach.}
        \label{fig:timmingComparison}
    \end{figure}

   For example, consider two nodes, A and B, both configured to check for instructions every 60 seconds. Suppose the current time is $t=0$; based on their prior wake-ups, Node A is expected to wake at $t=30$ and Node B at $t=40$. In order for this pair to perform ranging, the server needs to assign the task at a time later than the latest of the two wake-ups (i.e., $\geq 40$). Let us assume the server schedules both to perform ranging at $t=45$. In a traditional LoRaWAN approach illustrated in Fig. \ref{fig:timmingComparison}(a), the server might queue instructions and static time offsets, e.g., instructing Node A to do ranging in 15 seconds (i.e., $45-30$) and Node B in 5 seconds (i.e., $45-40$). If Node A experiences a collision at $t=30$ and retries at $t=32$, it would incorrectly interpret the original offset and execute at $t=47$ (i.e., $32+15$), breaking synchronization. By contrast, in our design illustrated in Fig. \ref{fig:timmingComparison}(b), the instruction is tied to the dynamic countdown rather than a static offset, so even after retransmission, Node A receives the correct remaining time (13 seconds at $t=32$), and both nodes execute the ranging at $t=45$ as intended. This mechanism makes the countdown strategy robust against collisions, prevents cumulative drift, and ensures that time-critical tasks remain properly coordinated across the network.
	
	\section{Operational Details of the Orchestration Mechanism}
	 
	In this section, we discuss the network server, which consists of several sub-systems. The NS is composed of four main components: an \textit{ MQTT broker}, a \textit{database}, a \textit{Task Manager}, and a \textit{Data Manager}. The MQTT broker and database can be implemented either internally or by using external services as long as they are appropriately configured to meet the requirements of the NS. 
	
	\subsection{MQTT broker}
	
	In this framework, we use the MQTT broker as a lightweight and efficient bridge between the GW and the NS. The MQTT broker acts as an intermediary that receives data from publishers and routes it to appropriate subscribers based on predefined topics. Rather than sending data directly from one device to another, devices communicate by publishing data to a central broker under a specific topic name. Subscribers interested in that topic then receive the data from the broker. This publish-subscribe pattern reduces the need for complex addressing and direct device management, enabling more scalable and decoupled system architectures. 
	
	In the proposed framework, when the GW receives data from a node, it forwards the data by publishing it to the broker, while the NS subscribes to the relevant topics to receive the transmitted data. Similarly, when the NS needs to send data to a node, it publishes the data to a designated topic, which the GW subscribes to, thereby enabling downstream communication. For instance, when an end node completes a ranging operation, it transmits the result to the GW. The GW inspects the received message and, based on the opcode indicating a ranging result, it forwards the data by publishing it to the \textit{ResultTopic}. On the NS side, the Data Manager, which is subscribed to the same topic, receives the published data and then proceeds with further processing. This communication scheme ensures modular, scalable, and asynchronous data exchange between components.
	
	\subsection{Database}

    The database at the NS provides the state needed to coordinate energy-aware behavior across nodes. It maintains three categories of information: (i) configuration records that define how each node operates; (ii) logs from instruction checks; and (iii) results of ranging and localization. Together, these records allow the scheduler to place countdown-based tasks that align with nodes’ next wake times without relying on a shared real-time clock.
    
    Each configuration record includes mandatory fields such as the node identifier, the instruction check interval, and the operating mode. The operating mode can be either \textit{always on}, where the node remains continuously active (suitable for anchors with abundant or continuous power), or \textit{low power}, where the node sleeps most of the time and wakes briefly to check for instructions or perform scheduled work. If a node is designated as an anchor, its latitude and longitude must also be included in the record. Importantly, configuration is not fixed: the server can enqueue updates to a node’s mode or interval, which are fetched and applied at the next instruction check, giving flexibility as application requirements evolve.
    
    Logs generated during instruction checks contain a timestamp, node and network identifiers, the battery level reported by the node, and the gateway-measured RSSI of the uplink. These logs are used by the scheduler to predict when each node will next wake. When a ranging task is requested, the scheduler derives the countdowns from this log information and delivers them at the nodes’ subsequent instruction checks, ensuring the intended peers become active simultaneously. In addition, the logs support monitoring: the battery level trend helps plan replacements, RSSI provides a coarse proxy of node-to-gateway distance, and the instruction check history confirms node liveness.
    
    Lastly, the database stores the outcomes of ranging. For point-to-point ranging, it records the master and slave identifiers, the frequency-error-assisted distance estimate, and the RSSI of the response. For passive ranging, it tracks which listener overheard a given master–slave exchange and stores the raw time-difference measurement. When a sufficient number of measurements are accumulated in a batch, the data can then be used (or is readily available) to compute a position estimate for the target.
    
    By unifying configuration, logs, and measurement results, the database enables synchronized wake scheduling, supports device monitoring, and provides a foundation for reproducible post-processing of localization data.
    
    \subsection{Task Manager}
    
    The task manager orchestrates node behavior using the information stored in the database and user requests received via the API. Communication between the NS and nodes follows a predefined frame structure (Fig.~\ref{fig:COmmunFrame}). Its main functionalities are implemented as handlers:
    
    \subsubsection{\textbf{Instruction check handler}}
    
    Processes incoming instruction check messages, logs the node’s battery level and RSSI, and checks for any pending tasks. If tasks exist, the handler assigns them with countdowns derived from the most recent instruction check logs, ensuring that peer nodes wake in sync for ranging.
    
    \subsubsection{\textbf{Configuration check handler}}
    
    Handles nodes that power on or require updated configuration. It retrieves the node’s parameters (mode, interval, and anchor location if applicable) from the database and delivers them. Configuration updates written into the database are automatically applied at the next instruction check.
    
    \subsubsection{\textbf{Ranging instruction handler}}
    
    This handler compiles multiple ranging instructions into a single batch. In a ranging operation, two nodes are involved; thus, a single ranging instruction results in a queued task for both participating nodes (master and slave nodes). Batching is particularly beneficial in localization scenarios, where multiple distance measurements (typically between a target node and several anchor nodes) are required to accurately estimate the target’s location.
    
    \begin{figure}[!t]
        \centering
        \includegraphics[width=0.65 \columnwidth]{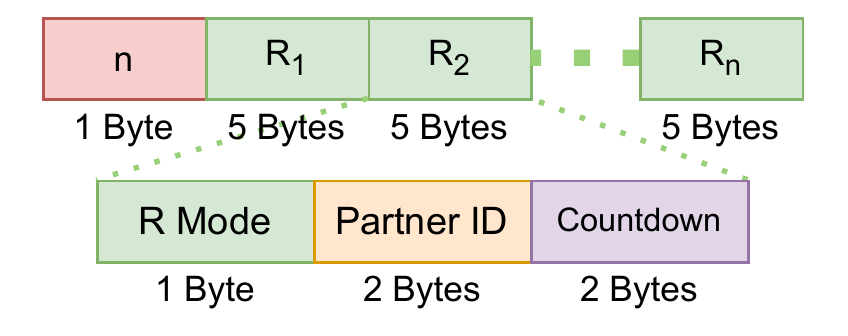}
        \caption{Structure of a batch point-to-point ranging instruction.}
        \label{fig:ranging_batch}
    \end{figure}
    
    Fig.~\ref{fig:ranging_batch} illustrates the structure of a batch ranging instruction. It consists of two main components: the total number of ranging tasks in the batch, denoted by $n$, and a corresponding set of ranging property entries, $R_i, \,i = 1, 2, ..., n$. Each $R_i$ entry contains three fields: \textit{mode}, \textit{partner ID}, and \textit{countdown}. The mode specifies whether the node acts as a master or a slave during the ranging exchange. The pair ID identifies the corresponding node with which distance should be measured, and this also implicitly defines the role of the node (i.e., the other node in the pair will take the opposite role), and the countdown represents the remaining time before the instruction must be executed.
    
    \subsubsection{\textbf{Passive ranging instruction handler}}
    
    This handler coordinates passive listeners that monitor master–slave exchanges. Each passive listener node specifies which link to observe and when, enabling time-difference measurements without additional transmissions.
    
    \begin{figure}[!ht]
        \centering
        \includegraphics[width=0.65 \columnwidth]{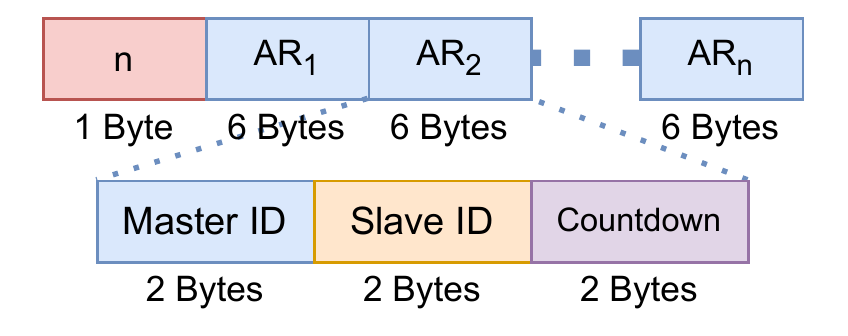}
        \caption{Structure of a batch passive ranging instruction.}
        \label{fig:ranging_batch_passive}
    \end{figure}
    
    Fig.~\ref{fig:ranging_batch_passive} shows the structure of a batch passive ranging instruction. It consists of two main components: the total number of passive ranging tasks in the batch, denoted by $n$, and a corresponding set of passive ranging property entries, $AR_i, \, i = 1, 2, ..., n$. Each $AR_i$ entry contains three fields: \textit{master ID}, \textit{slave ID}, and \textit{countdown}.
    
    \subsubsection{\textbf{User API}}
    
    Provides an interface for end users to modify configurations or request for ranging tasks. Commands received via HTTP or HTTPS are translated into tasks and queued for nodes. The API also enables integration with higher-layer applications such as dashboards or GUIs for monitoring and control.

	\subsection{Data Manager}
	
    The data manager is responsible for storing ranging results, which can originate from two sources: point-to-point ranging exchanges and passive ranging listeners. 
    
    In PtP ranging, which is conducted in batches, the master node generates a unique random ranging ID that tags all measurements within the batch. This identifier allows the data manager to group related results for a given batch localization task. Once a PtP ranging process is completed, the master node immediately sends the measurement results to the GW, which forwards them to the NS, where they are stored for further use. 
    
    In passive ranging, each batch is also assigned a random ranging ID. To avoid wireless transmission collisions, passive nodes do not transmit their results immediately after the master–slave exchange. Instead, each node generates a bounded random countdown timer that determines when its data will be sent. This staggered transmission strategy significantly reduces the likelihood of collisions. Once received at the NS, the passive ranging results are also stored. 
    
    After data from both PtP and passive ranging are collected, the most recent measurements associated with the same ranging ID can be processed to estimate the location of the target node.

	\section{Implementation, Experiment, and Results}
	
\begin{figure}[!t]
		\begin{subfigure}{0.49\linewidth}
			\includegraphics[width=\linewidth]{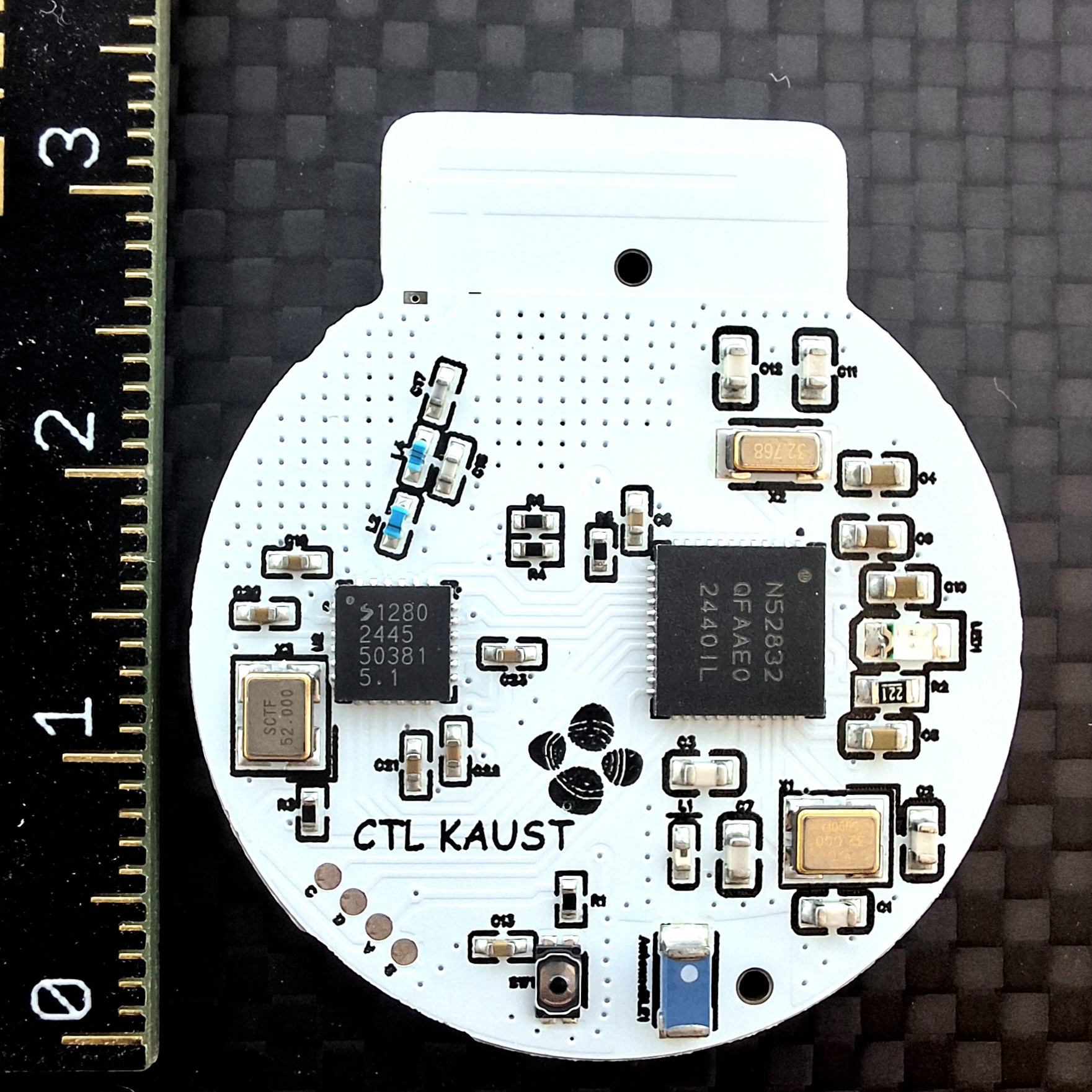}
			\caption{ }
		\end{subfigure}
		\begin{subfigure}{0.49\linewidth}
			\includegraphics[width=\linewidth]{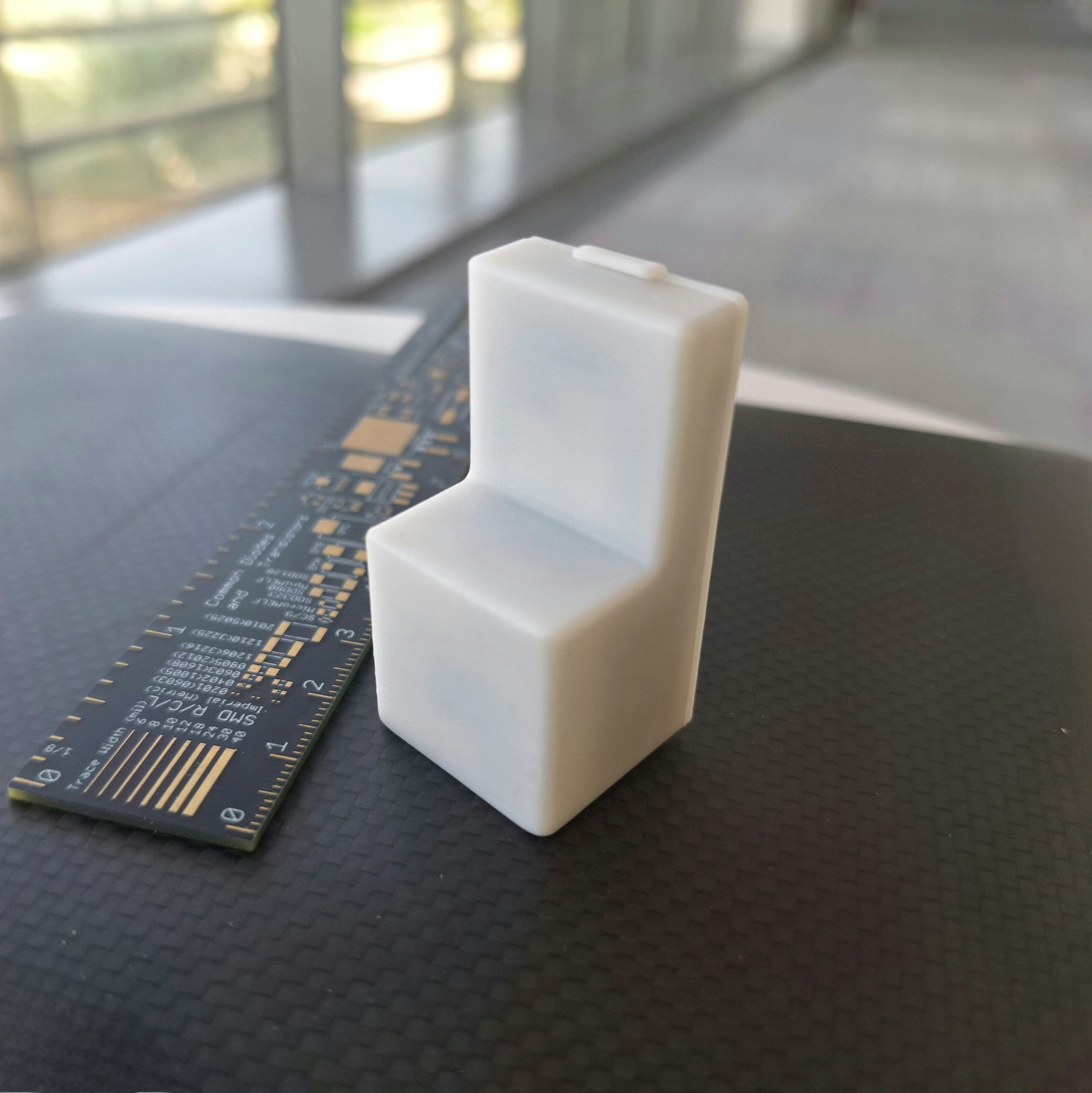}
			\caption{ }
		\end{subfigure}
		\caption{(a) The node; (b) The gateway}
		\label{fig:hardware}
	\end{figure}

	We validated the proposed framework by implementing the nodes, GW, and NS. We designed and developed a compact wireless node, incorporating an ARM Cortex M-based microcontroller, interfaced with an SX1280 LoRa transceiver. The gateway is also custom-built using an Xtensa-based microcontroller with TCP/IP capabilities, interfaced with an SX1280 chip. The hardware designs for both components are shown in Fig.~\ref{fig:hardware}.  Lastly, the NS was deployed as a containerized application, providing an isolated and portable environment that is easy to deploy on physical machines or cloud infrastructure.
	
\begin{figure}[htbp]
		\begin{subfigure}{0.693\linewidth}
			\includegraphics[width=\linewidth]{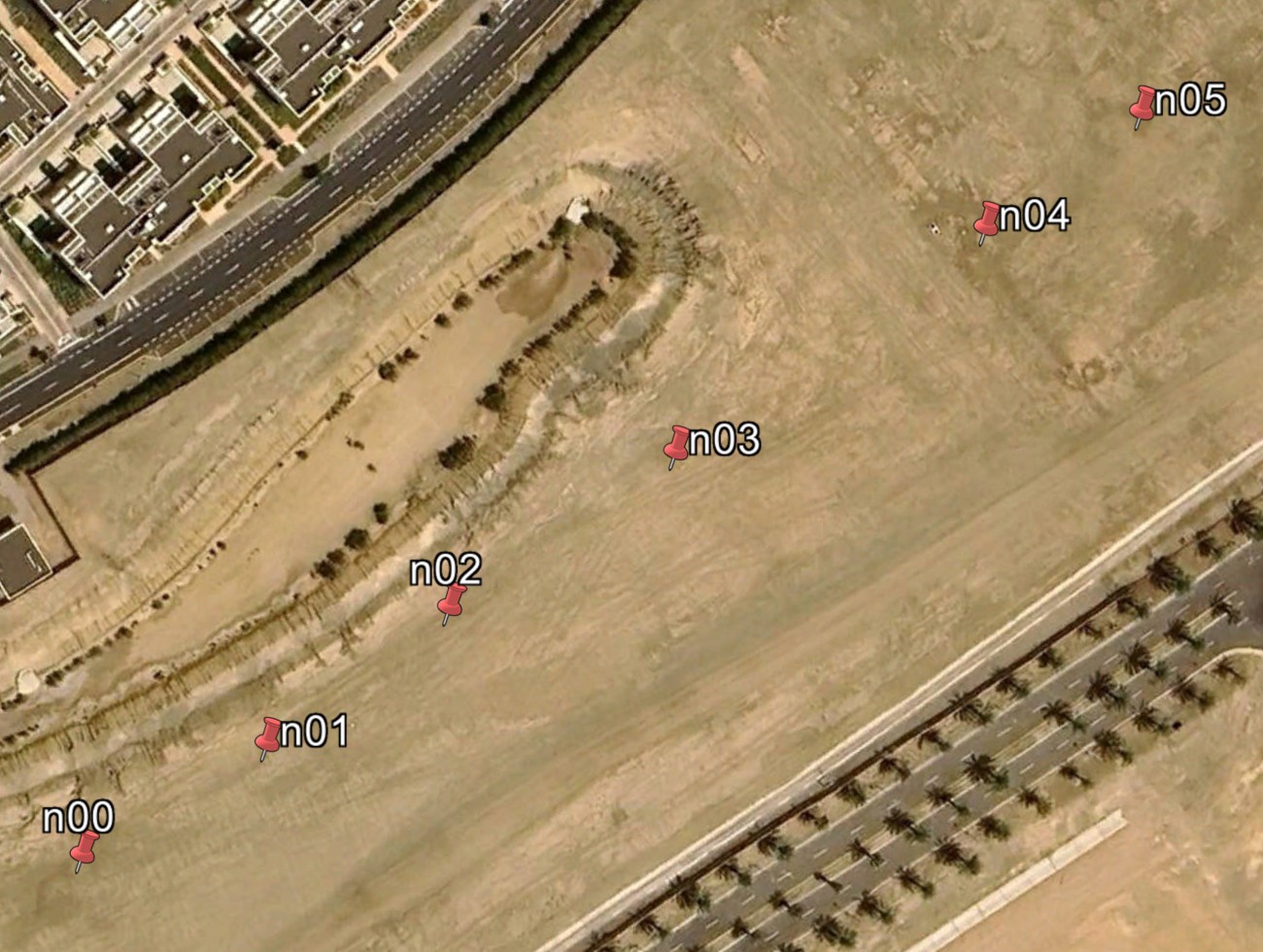}
			\caption{ }
		\end{subfigure}
		\begin{subfigure}{0.293\linewidth}
			\includegraphics[width=\linewidth]{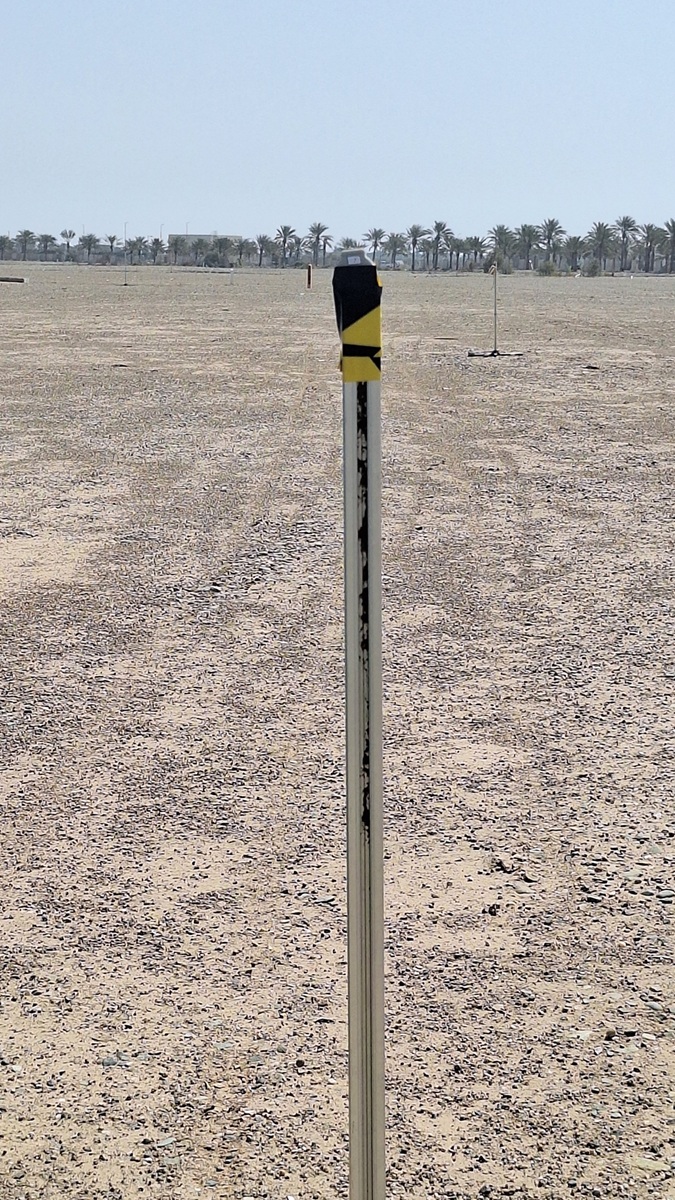}
			\caption{ }
		\end{subfigure}
		\caption{(a) Satellite image of the nodes location; (b) A single node on a pole}
		\label{fig:peta}
	\end{figure}

	Following system deployment, we conducted node calibration by placing the devices in an open field with minimal obstructions. Each node was mounted on a 1-meter-high pole, as illustrated in Fig.~\ref{fig:peta}. 

    \begin{figure}[!b]
        \centering
        \begin{subfigure}[b]{0.85\columnwidth}
            \centering
            \includegraphics[width=\linewidth]{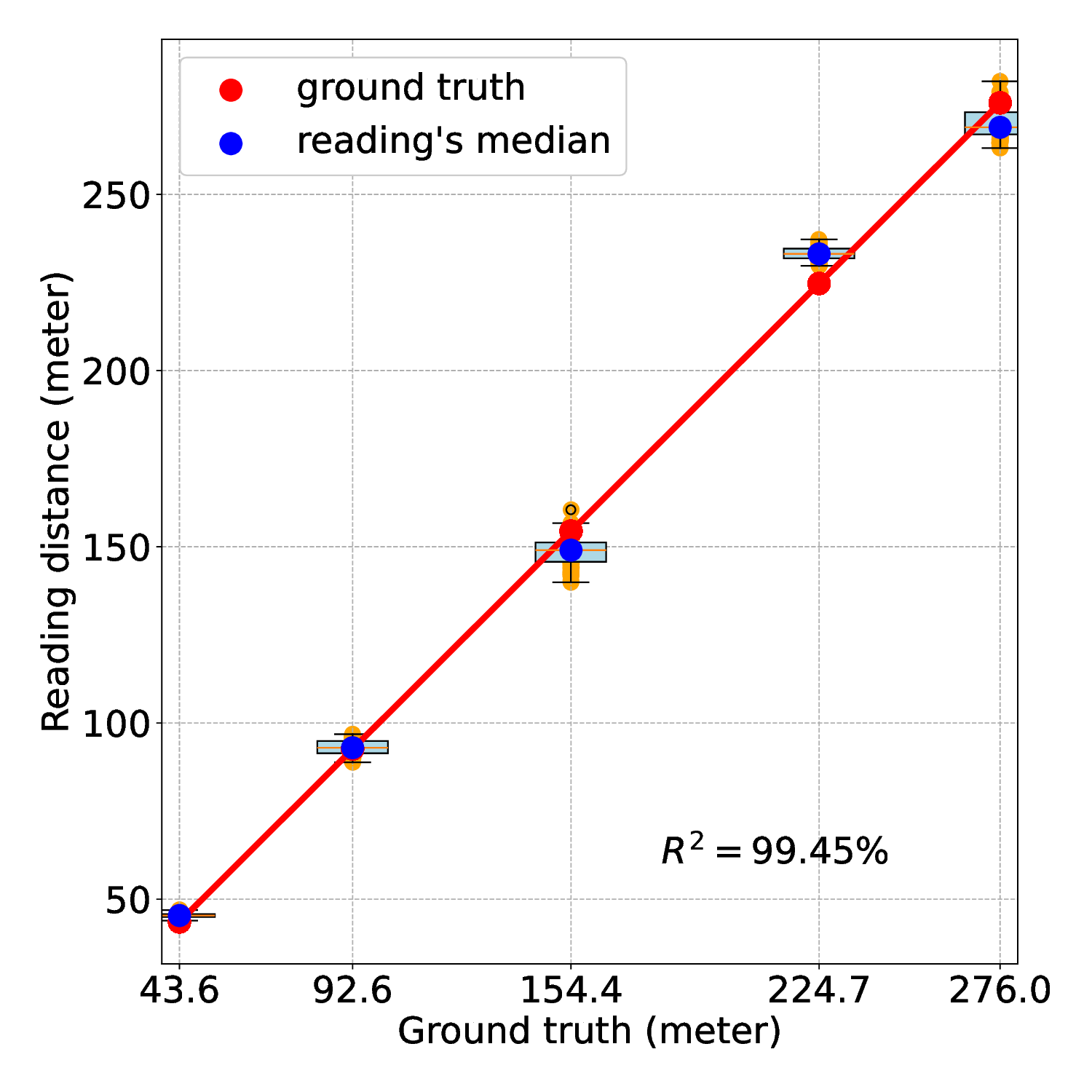}
            \caption{}
            \label{fig:scatterLinear}
        \end{subfigure}
        \hfill
        \begin{subfigure}[b]{0.85\columnwidth}
            \centering
            \includegraphics[width=\linewidth]{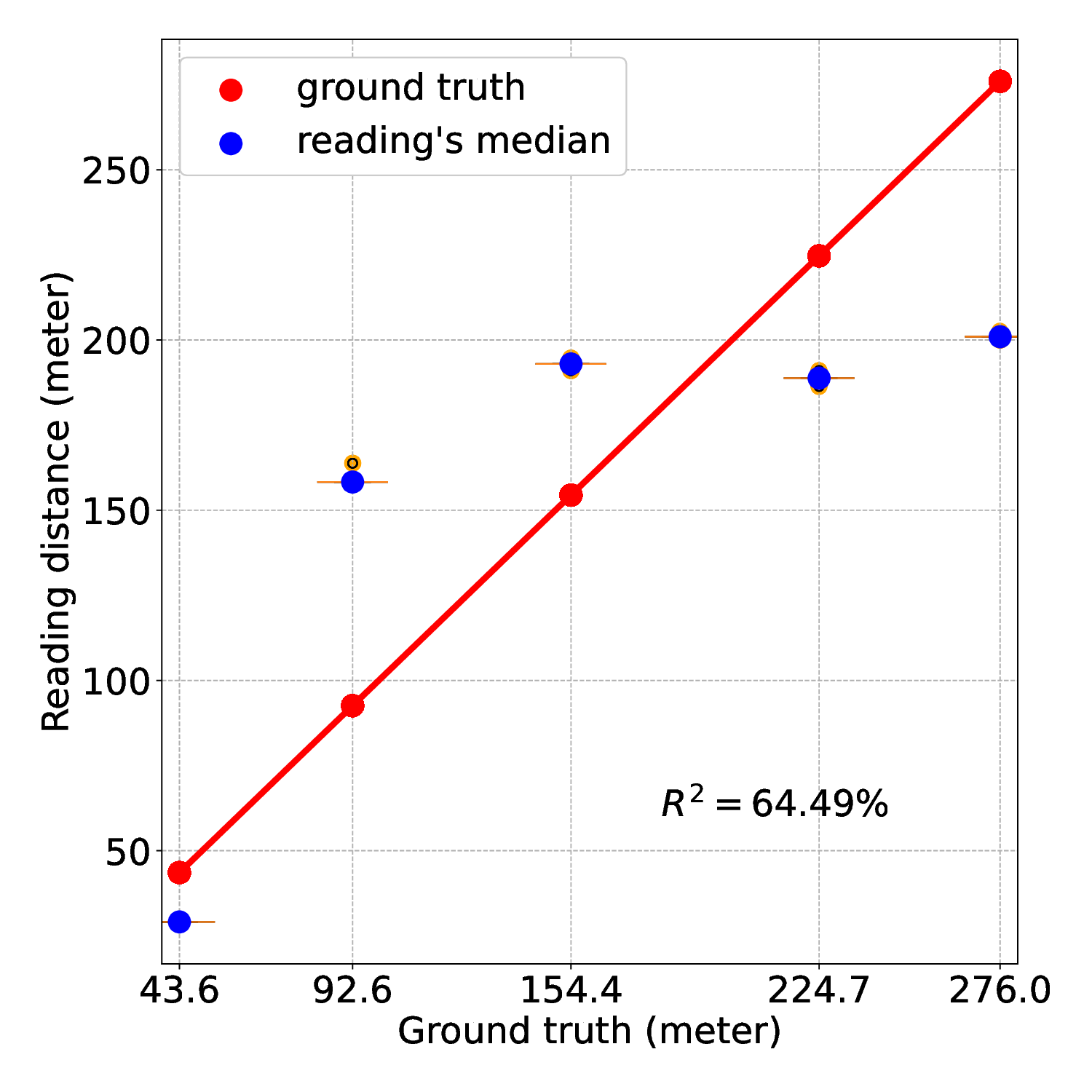}
            \caption{}
            \label{fig:scatterLinearrssi}
        \end{subfigure}
        
        \caption{Comparison between distance prediction methods: 
        (a) FEI-corrected TW-ToA, and 
        (b) RSSI-based.}
        \label{fig:scatterComparison}
    \end{figure}
	We used an RTK GPS module (Ublox Z9F) to record precise ground truth positions for each node, achieving centimeter-level accuracy. In the deployment, the nodes were configured with a bandwidth of 1625~kHz, SF10 for communication, SF8 for ranging, a transmit power of 12~dBm, and a 12-symbol preamble. To minimize interference between the two functions, communication and ranging were assigned different center frequencies. These specific configuration parameters were selected arbitrarily, as the proposed framework functions independently of these parameter values. Communication and ranging were conducted on separate channels. A 10 ppm crystal oscillator was used in the hardware design, which meets the SX1280’s requirement (maximum ±80 ppm at SF8, 1625 kHz).
	
	Calibration was carried out by instructing each node to perform 10 ranging operations with a designated reference node ($n00$). For each session, raw ranging values, Frequency Error Information (FEI), and ranging RSSI were recorded and subsequently averaged to obtain stable measurements.

	Fig.~\ref{fig:scatterComparison}(a) presents distance predictions derived from TW-ToA-based measurements obtained using linear regression on the collected data. A total of 70 ranging samples were collected per node relative to node $n00$. While some deviations are present, the data exhibit strong linearity, with an $R^2$ value of approximately 99.45\%, indicating highly consistent ranging performance. The median of the estimated positions (blue dot) closely aligns with the ground truth location (red dot), confirming that the LoRa ranging engine maintains its high precision. 
	For comparison, the RSSI-based approach is also evaluated and is observed to exhibit lower linearity, as shown in Fig.~\ref{fig:scatterComparison}(b), It exhibits a significantly lower $R^2$ value compared to the TW-ToA-based method. This highlights the superior precision of the TW-ToA approach over the RSSI-based alternative.

    \begin{figure}[!t]
		\centering
		\includegraphics[width=0.85\columnwidth]{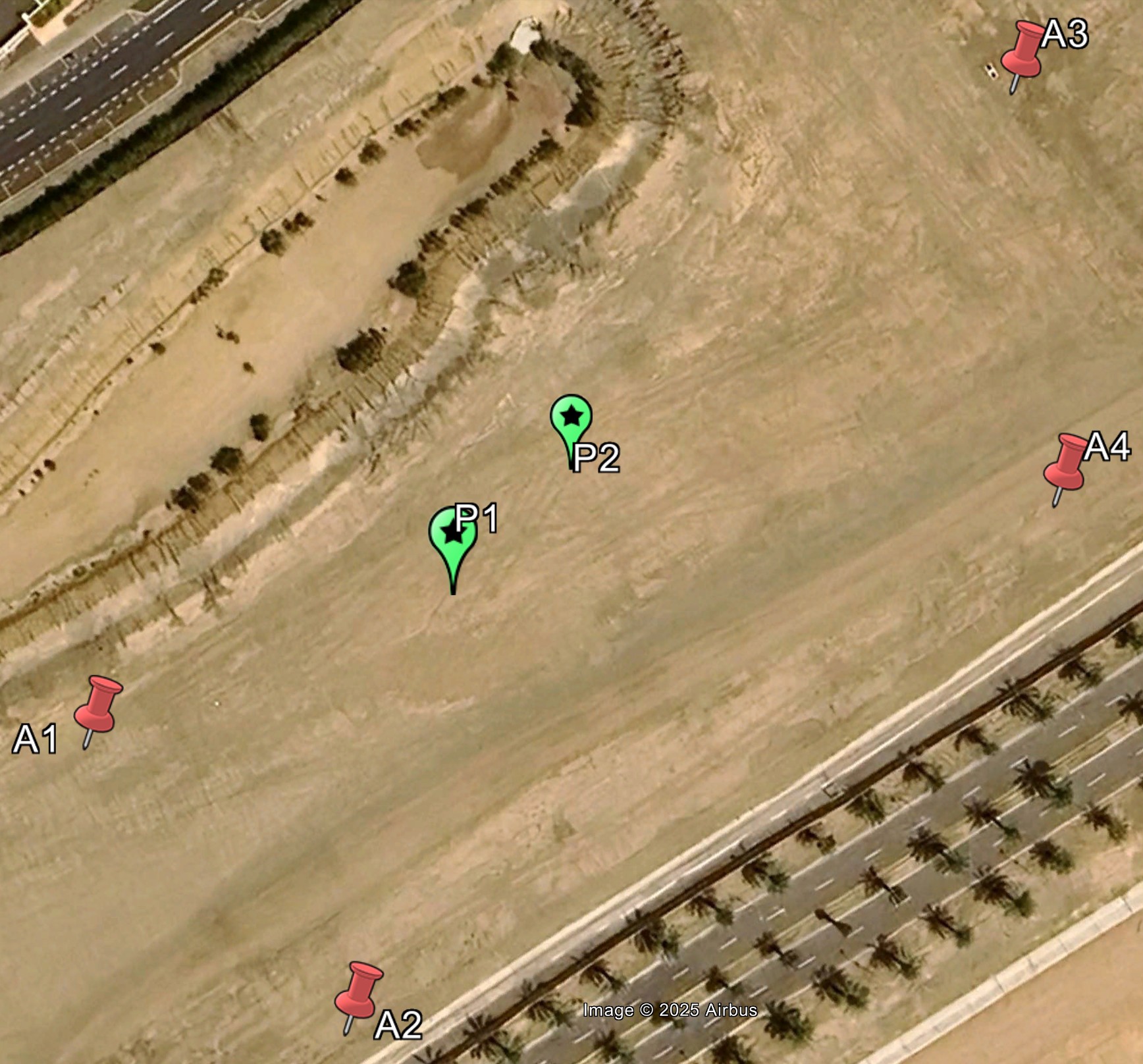}
		\caption{Node placement for the localization experiment.}
		\label{fig:map_ptp}
	\end{figure}
	
	\subsection{Localization Experiment}
	
	Following the calibration process, the resulting parameters are used to convert raw ranging measurements into accurate distance estimates. These estimates form the basis for a subsequent localization experiment, in which four anchor nodes and two target nodes are deployed over an area of approximately 180~m $\times$ 180~m, as illustrated in Fig.~\ref{fig:map_ptp}.
	
	\begin{figure}[!t]
		\centering
		\includegraphics[trim=50 0 65 0, clip, width=0.9\columnwidth]{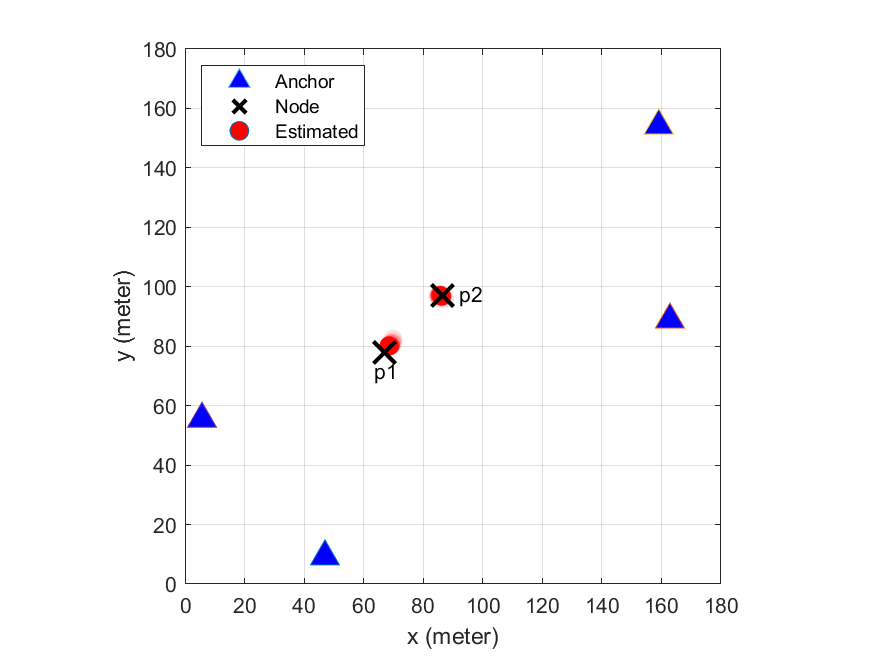}
		\caption*{(a)}
		
		\begin{minipage}[b]{0.45\columnwidth}
			\centering
			\includegraphics[trim=50 0 80 0, clip, width=\columnwidth]{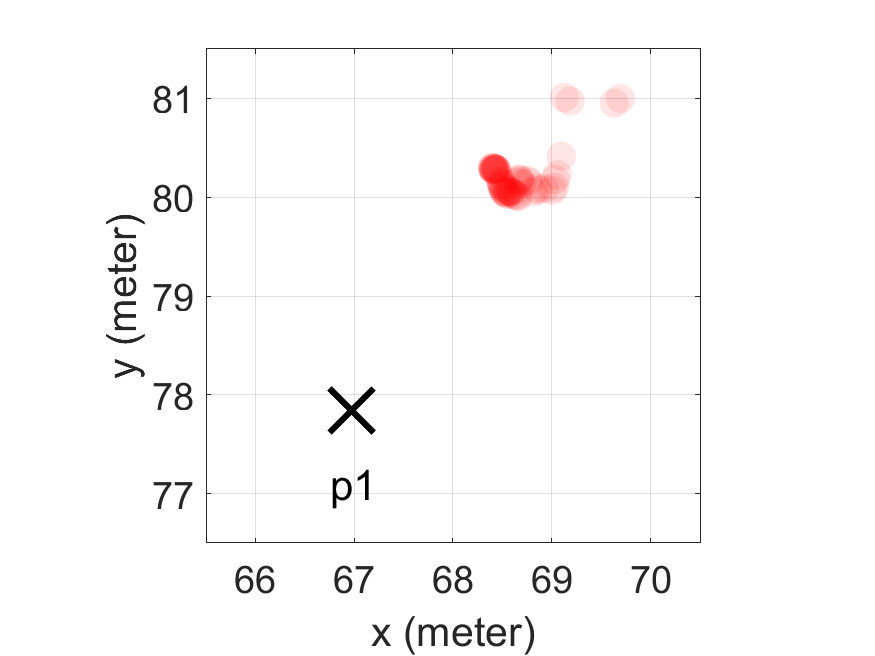}
			\caption*{(b)}
		\end{minipage}
		\begin{minipage}[b]{0.4223\columnwidth}
			\centering
			\includegraphics[trim=70 0 80 0, clip, width=\columnwidth]{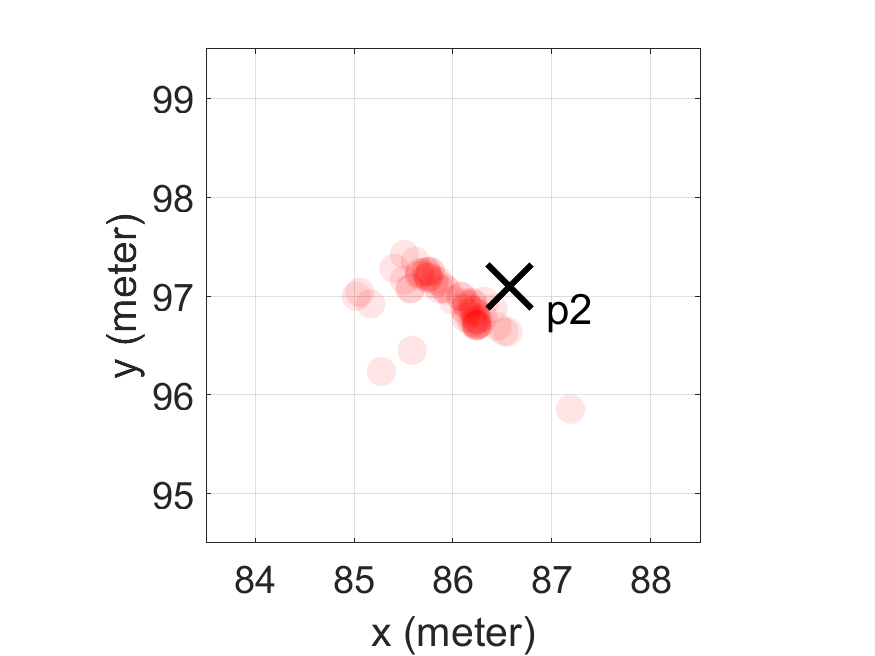}
			\caption*{(c)}
		\end{minipage}
		
		\caption{Localization results: (a) overall node placement and estimated targets' positions; (b) and (c) zoomed-in localization results for the first and second target nodes, respectively.}
		\label{fig:LocResults}
	\end{figure}
	
	Fig.~\ref{fig:LocResults} presents the localization results obtained using a least squares estimator. Each point in the experiment is based on 60 individual measurements with respect to all anchors, which are automatically processed by our proposed framework. As expected, the SX1280 chip demonstrated reliable ranging performance, resulting in precise position estimations. The Root Mean Square Error (RMSE) for the first target is 3.0~m, while the second target achieves an RMSE of 0.8~m. These results indicate that even when the device is activated for only a few milliseconds during the ranging process, the localization performance remains unaffected.  

    We emphasize that the localization algorithm used here is intentionally simple, since the primary focus of this work is on energy-aware scheduling rather than advanced solver design or NLoS mitigation. The proposed framework is solver-agnostic and can readily be combined with more sophisticated algorithms.
	
	\subsection{Energy Consumption Analysis}
	
	Efficient power management is crucial in wireless communication systems, particularly within the context of the proposed scenario. This section analyzes the device’s energy consumption characteristics under a duty-cycled operation scheme. The device predominantly remains in sleep mode, waking at fixed intervals for instruction checks or at scheduled times to perform specific tasks like ranging. To estimate power usage, we measure current consumption under a constant voltage of 3 volts using the Power Profiler Kit II from Nordic Semiconductor.
	
	The operational cycle of the node is divided into four key processes: \textit{instruction check request}, \textit{instruction check response}, \textit{ranging}, and \textit{sleeping}. During the instruction check request process, the end node transmits a payload to the gateway to inquire whether any tasks are pending. In the subsequent instruction check response process, the node listens for instructions from the gateway. Here, the node employs Channel Activity Detection (CAD) to detect the presence of a preamble to minimize energy consumption during this process. Specifically, the node is configured to detect a preamble with 8 symbols to reduce the likelihood of false positives, which could incorrectly indicate that an instruction is incoming. If no preamble is detected, it assumes no instructions are being sent and skips opening the receive window, thereby conserving energy.
	The third process is the ranging process, where the node performs a configurable number of ranging repetitions and sends the averaged result to the gateway. The last is the sleeping process, which occupies the majority of the cycle. This phase is designed to minimize energy usage, as the device remains idle for an extended period. A summary of the energy consumption across each of these stages within a single operational cycle is provided in Table~\ref{table_power}.

	\begin{table}[!h]
		\centering
		\caption{Energy consumption in one cycle by the node, measured at a 3-volt supply}
		\label{table_power}
		\begin{tabular}{|c|c|c|c|c|}
			\hline
			\textbf{Process Name} & \textbf{Avg. Current } & \textbf{Dura}- & $Q$ & $E$ \\
			   & \textbf{Consumption} & \textbf{tion}  & (mC) & (mJ)\\
			\hline
			Instruction check req. & 23.5 mA & 41 ms & 0.9635 & 2.8905 \\ 
			\hline
			Instruction check resp.   & 12 mA  & 15 ms & 0.18 & 0.54\\ 
			with only CAD &   &  & & \\ 
			(no received packet)& & & & \\
			\hline
			Instruction check resp. & 15 mA  & 106 ms  & 1.59 & 4.77\\ 
			(with 8 bytes of  & &   &  & \\ 
			dummy packet) & & &  &  \\
			\hline
			PtP ranging with 10& 22 mA & 379 ms & 8.338  & 25.01\\ 
			repeats + send result & & &  & \\
			\hline
			Sleeping & 2.5 $\mu$A & - & -  & -\\ 
			\hline
		\end{tabular}
	\end{table}

    To conveniently analyze battery endurance, we first use the electric charge notation $Q$, measured in coulombs (C):
    \begin{equation}
        Q = I \times t,
    \end{equation}
    where $I$ is current in amperes and $t$ is the operation duration in seconds. This representation is directly comparable to the rated charge of a battery. For completeness, we also report the equivalent energy consumption, given by
    \begin{equation}
        E = V \times Q,
    \end{equation}
    where $V$ is the supply voltage.
    
    Let us consider a scenario in which we aim to evaluate the endurance of a device powered by a single battery. Each operational cycle consists of a pair of instruction check request and response, during which no actual instruction is received from the gateway, followed by a sleep phase for the rest of the cycle. Each cycle has a total duration of $\tau$ seconds. Our objective is to estimate how long the battery will last under these conditions. We define $Q_T^{(\tau)}$ as the total electric charge consumed during a single cycle. Based on the measured consumption of each activity, the total charge for a cycle can be expressed as:
	\begin{equation}
		Q_{T}^{(\tau)} = 0.96+0.18+2.5\cdot10^{-3}\cdot(\tau-56\cdot10^{-3})
	\end{equation}

	As a practical example, suppose we set the cycle duration to 10 minutes ($\tau = 600$). Additionally, let us assume the device is powered by a CR2032 battery, which typically has a nominal capacity of 225 mAh, equivalent to 810 C. This setup allows us to calculate how many cycles $f_{cyc}^{(\tau)}$ the battery can support by:
	\begin{equation}
		\begin{aligned}
			f_{cyc}^{(\tau)}&=\frac{Q_{B}}{Q_{T}^{(\tau)}}
			=\frac{810\cdot10^{3}}{2.6434}
			=306,427 \; \text{cycles},
		\end{aligned}
		\label{eq:f_cyc}
	\end{equation}
	where $Q_B$ is the electric charge of the battery. Consequently, we can also estimate the device’s operational lifetime $t_{E}^{(\tau)}$, which can be expressed as:
	\begin{equation}
		\begin{aligned}
			t_E^{(\tau)}=f_{cyc}^{(\tau)}\cdot \tau
			=5.83 \; \text{years}.
		\end{aligned}
		\label{eq:t_E}
	\end{equation}
	However, it is only in an ideal world where we can use 100\% of the battery’s nominal capacity. In reality, several factors influence actual battery endurance. For example, the CR2032 coin cell, while capable of short bursts (typically \textless 1 second) up to 15 mA, is constrained by its design, including a low continuous current tolerance up to ~5 mA, high internal resistance, and load-induced voltage sag. Repeated high-current pulses beyond its capability led to voltage instability, reduced efficiency, and accelerated capacity degradation. As a result, the practical capacity under dynamic loads often falls far below the nominal capacity. Hence, we conduct real-world testing to evaluate the actual endurance of our device when powered by a single CR2032 battery, accounting for these practical factors.
	
	In the experiment, we initially considered a 10-minute cycle as the intended operational interval. However, conducting the test with such intervals would require a significant amount of time to gather meaningful data. To address this, we scale the cycle duration down to 30 seconds, allowing for quicker data collection. The results from this shorter cycle are then extrapolated to represent the 10-minute scenario. As more frequent activity bursts tend to degrade battery performance more severely than less frequent ones, the extrapolated result gives us a lower bound on the expected operational lifetime.
	
\begin{figure}[ht!]
		\centering
		\includegraphics[width=0.93\columnwidth, trim={7em 3em 8.5em 6.5em}, clip]{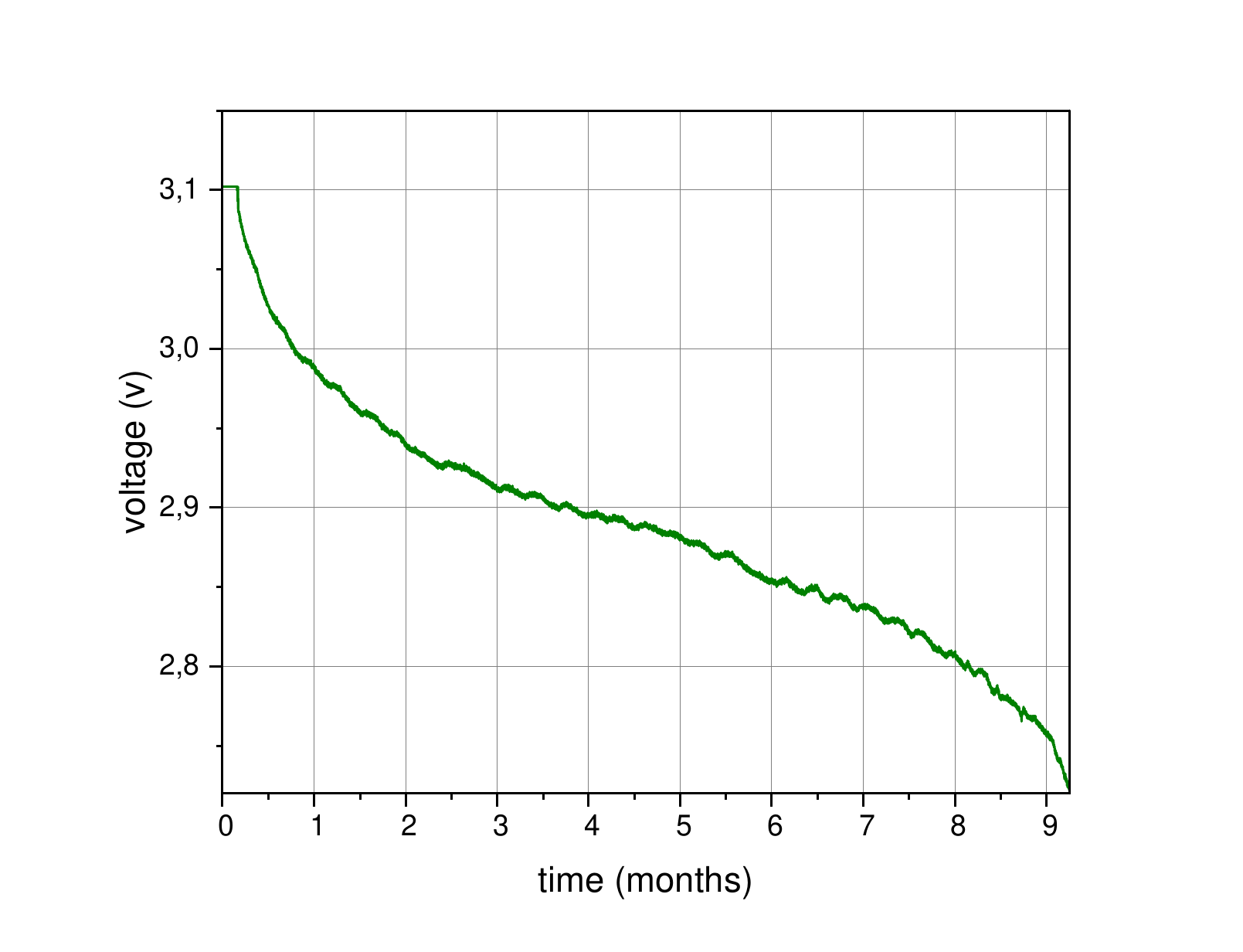}
		\caption{Battery voltage over time for a scenario with an instruction check routine in each cycle (10-minute cycle)}
        \vspace{-1em}
		\label{fig:power_instruction_check}
	\end{figure}

	In the first experiment, we used a scenario where each cycle consisted of an instruction check request and response, with no instruction sent from the gateway, followed by a sleep phase. For this test, the device was powered by a CR2032 battery, where we defined 3.1~V as the full capacity level (100\%) and 2.72~V as the end-of-life threshold (0\%). Under these conditions, the device operated for $t_E^{(30)} = 30.16$ days, completing a total of 86,867 cycles. From this result, we can estimate the  battery’s electric charge, denoted as $Q_B^{(prac)}$, as follows:
	\begin{equation}
		\begin{aligned}
			Q_B^{(prac)}=Q_T^{(30)}\cdot 86867
			=105.83 \; \text{C}.
		\end{aligned}
		\label{eq:Q_B_prac}
	\end{equation}
	Using this battery charge derived from the 30-second cycle test, along with equations (\ref{eq:f_cyc}) and (\ref{eq:t_E}), we can extrapolate the device’s expected operational lifetime under a 10-minute cycle scenario, denoted as $t_E^{(600)}$, by:
	\begin{equation}
		\begin{aligned}
			t_E^{(600)}=\frac{Q_B^{(prac)}}{Q_{T}^{(600)}}\cdot 600
			=9.26 \; \text{months}.
		\end{aligned}
		\label{eq:t_E_firstExp}
	\end{equation}
	Fig.~\ref{fig:power_instruction_check} illustrates the voltage level of the CR2032 battery over time, extrapolated to represent a 10-minute cycle scenario. As shown, the battery voltage gradually declines, reaching the defined end-of-life threshold of 2.72 V at approximately 9.26 months, which serves as a lower-bound estimate of the device’s endurance.
	
	For the next experiment, we used a scenario where each cycle consisted of an instruction check request and response, with a ranging instruction (involving 10 repetitions) sent from the gateway, followed by a sleep phase. The node executes the ranging process based on a countdown specified in the instruction and transmits the result back to the gateway; during the remaining time, it remains in sleep mode. As in the previous experiment, we used a 30-second cycle for testing.

    According to the results, the node operated for $t_E^{(30)} = 2.74$ days, completing a total of 7,891 cycles. From this, we estimate the battery electric charge using \eqref{eq:Q_B_prac} as $Q_B^{(30)} = 75.41\; \text{C}$. This value is significantly lower than in the first experiment, primarily due to more frequent discharge bursts and the relatively long duration of the ranging process, which lasts approximately 379 ms and exceeds the typical peak current tolerance of the CR2032 (around 15 mA for bursts shorter than 1 second). Similar to the previous case, we extrapolate the expected operational lifetime under a 10-minute cycle using the obtained $Q_B^{(30)}$. Applying the same method as in \eqref{eq:t_E_firstExp}, the node is projected to last for $t_E^{(600)} = 47.6$ days.

    \begin{figure}[ht!]
    	\centering
    	\includegraphics[width=0.93\columnwidth, trim={7em 3em 8.5em 6.5em}, clip]{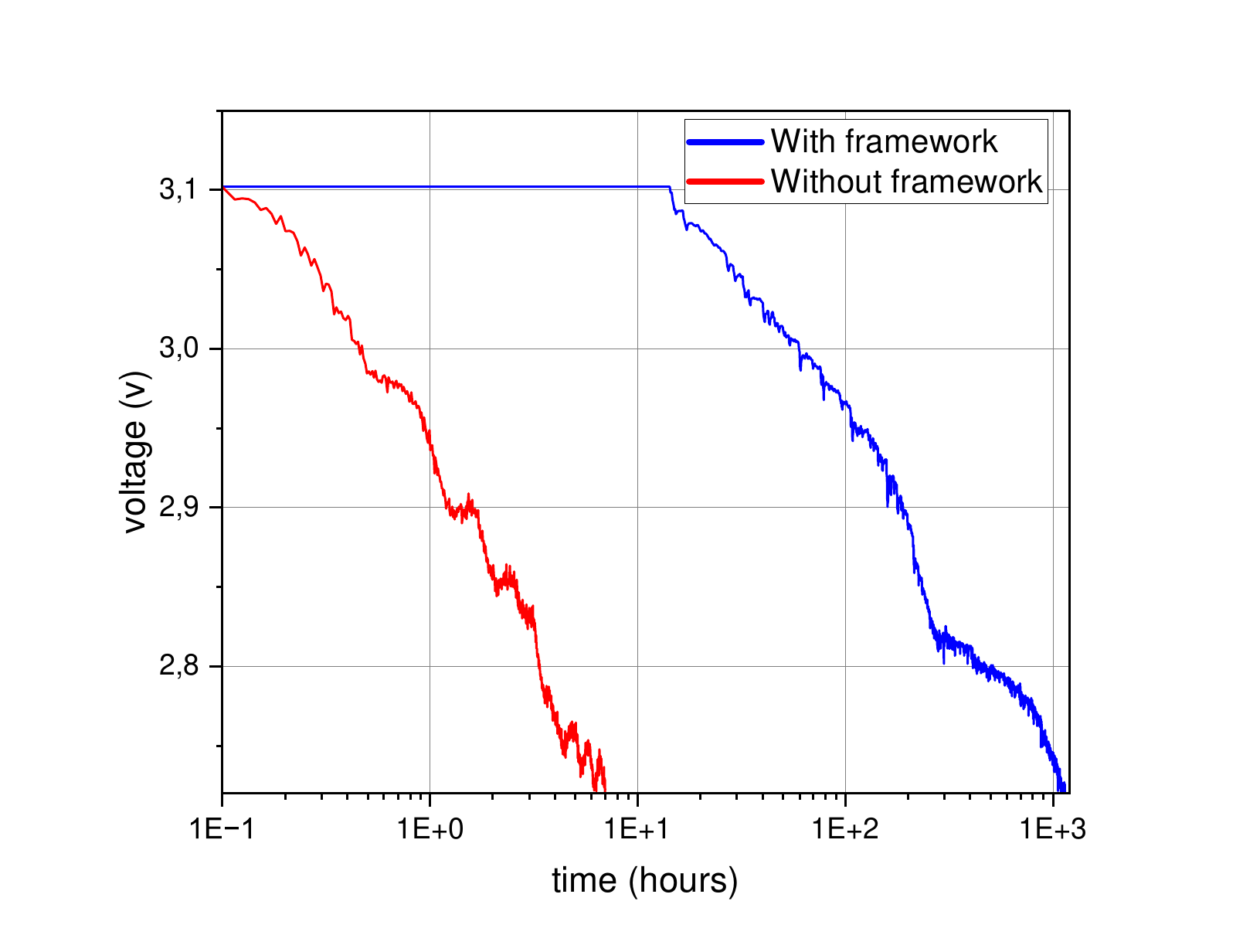}
    	\caption{Battery voltage over time with and without the proposed framework for a single ranging operation per cycle (10-minute cycle).}
        \vspace{-1em}
    	\label{fig:framework_comparison}
    \end{figure}
    
    For comparison, we also examined baseline configurations without the proposed orchestration framework. The first baseline keeps nodes in continuous listening mode so that any ranging request can be detected immediately. While conceptually simple, this design consumes a continuous current of about 22 mA and drains a CR2032 cell in under 10 minutes, making it impractical for IoT deployments. The second baseline uses a CAD-assisted scheme, where each node scans the channel every 100 ms and opens a short receive window upon detecting channel activity. To enable ranging in this setup, the initiator must transmit a request with a long preamble ($\geq$100 ms) to ensure detection by the peer. Testing this configuration with a 30-second cycle (extrapolated to 10 minutes) resulted in only $t_E^{(600)} = 6.96$ hours of projected lifetime.
    
    These results, summarized in Fig.~\ref{fig:framework_comparison}, highlight the advantage of the proposed framework. Without orchestration, both always-on and CAD-assisted approaches lead to rapid energy depletion and poor scalability. In contrast, our framework coordinates wake-ups centrally, avoiding unnecessary channel scanning and enabling significantly longer lifetime on a CR2032. Under the same task of performing ranging every 10 minutes, the proposed framework exhibits an operational lifetime of 47.6 days (1142 hours) compared to just 6.96 hours for the CAD-assisted scheme, an improvement of more than two orders of magnitude ($\sim 160\times$).

    Beyond configuration flexibility, several hardware-level optimizations can also improve energy efficiency. In this work, we selected an ARM Cortex-M4-based microcontroller, nRF52832, as the host for the end node. Many other microcontrollers with lower energy consumption could equally be interfaced with the LoRa ranging-capable transceivers. Likewise, battery performance may be improved by opting for cells with higher capacity or lower internal resistance, which better tolerate high burst currents. Overall, while hardware choices can significantly influence energy performance, the energy-saving framework we propose ensures that unnecessary node activations are minimized through orchestration, helping to extend device longevity regardless of the specific hardware used.

    \section{Future Work and Extensions}

    Building on the current framework’s focus on energy-aware orchestration, several avenues remain open for future research and development:
    \begin{itemize}
        \item Security mechanisms are not yet incorporated. Adding lightweight cryptography (e.g., session keys as in LoRaWAN ABP/OTAA) would help protect against man-in-the-middle attacks, and recent advances in ultra-light encryption could be adapted for low-power IoT.
        
        \item Beyond periodic scheduling, the framework could be extended with event-driven triggers for urgent updates. For example, an ultra-low-power IMU or environmental sensor could initiate ranging or reporting in response to motion, vibration, or pressure changes, complementing the countdown-based core of the design.
        
        \item AI-assisted prediction is another promising direction. Learning-based models could optimize wake-up intervals or forecast network traffic, improving slot allocation and responsiveness without modifying the underlying protocol.
        
        \item The framework is frequency-agnostic. Future implementations could exploit sub-GHz LoRa bands for longer range or combine terrestrial IoT with LPWAN-based satellite links for global coverage. Such heterogeneous deployments would raise new challenges in routing, coordination, and scalability.
        
        \item Finally, integrating sensing and communication (ISAC) within the orchestration layer could broaden applicability. For example, variations in the radio signals already exchanged for ranging could be analyzed to detect basic motion or environmental changes, allowing the system to provide communication, localization, and sensing together with minimal energy overhead.
    \end{itemize}
    
    These extensions would enhance robustness, adaptability, and scalability, making the framework suitable for increasingly diverse IoT deployments.

	\section{Conclusion}
	In this paper, we presented a localization framework based on the LoRa ranging-capable transceiver, designed to prioritize low power consumption and low cost with high precision. The system demonstrates strong potential as an efficient alternative to conventional localization solutions that typically demand more energy or higher costs. This framework is implemented and tested on custom-designed hardware that is compatible with the framework's requirements. By leveraging the framework's wake-up orchestration, a node can achieve ultra-low power consumption while maintaining sub-5-meter localization accuracy. With a 10-minute instruction check interval, the node can remain in standby for up to nine months on a single CR2032 coin cell, while still supporting on-demand ranging when triggered. The proposed design also addresses timing drift and coordination challenges through a centralized scheduling mechanism managed by the network server. This architecture enables scalable, battery-efficient deployments that are well-suited for real-world applications such as logistics tracking, agriculture, and smart city infrastructure, where extended battery life, reliable accuracy, and cost-effectiveness are critical. Beyond localization, this orchestration mechanism can be adapted for other network-driven tasks, such as on-demand sensing.

	


	\bibliographystyle{IEEEtran}
	\bibliography{IEEEabrv,bibliography_ex}
	
	
	

\end{document}